\documentclass[12pt]{article}   

\usepackage{float}
\usepackage{amsmath,amssymb,amsfonts,bm,amsthm}
\usepackage{enumerate}
\usepackage{epsfig}
\usepackage{graphicx,rotating,lscape}
\usepackage{appendix}
\usepackage{hyperref}
\usepackage{natbib} 
\usepackage{booktabs} 
\usepackage[all]{xy}
\usepackage{color}
\usepackage{latexsym}
\usepackage{multirow}
\usepackage{graphicx}
\newtheorem{theorem}{Theorem}      

\newtheorem{definition}{Definition}

\newtheorem{corollary}{Corollary}
\newtheorem{remark}{Remark}
\theoremstyle{definition}
\newtheorem{example}{Example}
\newtheorem{Case}{Case Study}

\title{Model diagnostics of discrete data regression: a unifying framework using functional residuals}
\vspace{-0.4in}
\author{Zewei Lin and Dungang Liu\footnote{Zewei Lin is Ph.D. student (\href{mailto:linzw@mail.uc.edu}{linzw@mail.uc.edu}) and Dungang Liu is Associate Professor (\href{mailto:liudg@ucmail.uc.edu}{dungang.liu@uc.edu}), Department of Operations, Business Analytics and Information Systems, University of Cincinnati Lindner College of Business, Cincinnati, Ohio, USA.}}
\date{\today}

\begin{document}
\renewcommand{\arraystretch}{1.1}
\baselineskip = 8.1mm
\parskip = 4mm

\thispagestyle{empty}
\maketitle

\newpage
\setcounter{page}{1} 
\begin{center}
	\LARGE
Model diagnostics of discrete data regression: a unifying framework using functional residuals
\end{center}
  \vspace{1cm}
  \begin{abstract}
  \baselineskip = 6mm
   Model diagnostics is an indispensable component of regression analysis, yet it is not well addressed in standard textbooks on generalized linear models. The lack of exposition is attributed to the fact that when outcome data are discrete, classical methods (e.g., Pearson/deviance residual analysis and goodness-of-fit tests) have limited utility in model diagnostics and treatment. This paper establishes a novel framework for model diagnostics of discrete data regression. Unlike the  literature defining a single-valued quantity as the residual, we propose to use a function as a vehicle to retain the residual information. In the presence of discreteness, we show that such a functional residual is appropriate for summarizing the residual randomness that cannot be captured by the structural part of the model. We establish its theoretical properties, which leads to the innovation of new diagnostic tools including the functional-residual-{\it vs}-covariate plot and Function-to-Function ({\it Fn-Fn}) plot. Our numerical studies demonstrate that the use of these tools can reveal a variety of model misspecifications, such as not properly including a higher-order term, an explanatory variable, an interaction effect, a dispersion parameter, or a zero-inflation component. The functional residual yields, as a byproduct, Liu-Zhang's  surrogate residual mainly developed for cumulative link models for ordinal data (Liu and Zhang, 2018, {\it JASA}). As a general notion, it considerably broadens the diagnostic scope as it applies to virtually all parametric models for binary, ordinal and count data, all in a unified diagnostic scheme. 
   
   \vspace{0.3in}
	
\end{abstract}
\newpage
\section{Introduction}
In statistical modeling of business data, model diagnostics is an indispensable component. It helps domain researchers reassess their working models which are used to inform business decisions and strategies. However, except for a small class of linear regression models, model diagnostics have not been well addressed in business analytics and statistics in general. This lack of exposition is attributed to the fact that consumer or corporate data are often discrete. They could be binary, ordinal, or integer values. For example, binary data are used to indicate the bankruptcy/default status or disease prevention/treatment outcome; ordinal data are commonly seen in ratings of bonds, school districts, and pain severity; and integer-valued count data are prevalent in (electronic) records of insurance claims, emergency room visits, and frequency of product/device usage \citep{franses2001quantitative, de2008generalized, frees2009regression, stokes2012categorical, tutz2012regression, faraway2016extending}. When outcome data are discrete, it is more appropriate to use generalized linear models or other non-linear models to carry out statistical inference and inform business decisions. But for these non-linear models, classical methods (e.g., Pearson/deviance residual analysis and goodness-of-fit tests) have limited utility in model {\it diagnostics and treatment}. The lack of an effective tool results in an elevated  risk of model misspecification and unmeasurable bias in inference. In this paper, we establish a novel framework for {\it model diagnostics} of discrete data regression, with the goal of generating actionable insights for {\it model treatment}. 

Regardless of the types of discrete data, traditional goodness-of-fit assessments of a regression model heavily rely on hypothesis testing \citep{pyne1979single, lipsitz1996goodness, hosmer1997comparison, wang1997modeling, archer2006goodness, ARCHER20074450, blochlinger2011new, fagerland2016tests, fernandez2016goodness,fernandez2020model}. Although technical details may  vary from a specific model/setting to another, a hypothesis testing procedure ends up with a  $p$-value, which merely leads to a dichotomous decision -- rejection of the working model or not. In recent years, however, research and business communities are increasingly concerned with the use of $p$-values in decision-making \citep{simmons2011false, lambdin2012significance, nuzzo2014statistical}. In the article {\it ASA Statement on Statistical Significance and $P$-Values}, it was  stated that one of the principles we should uphold is ``Scientific conclusions and business or policy decisions should not be based only on whether a $p$-value passes a specific threshold'' \citep{wasserstein2016asa}. In the context of model assessments, if we agree that ``All models are wrong, but some are useful'' \citep{box1976science} as an acknowledgement of the complexities of reality, what do we expect a $p$-value to tell us? A small $p$ value may simply be a result of the sample size being large, which has become common in electronically recorded data sources across research disciplines and business industries; on the other hand, a large $p$ value may be an indication that the data set is not large enough to provide the ``resolution'' so that we can see the difference between the working and the true models \citep{wasserstein2016asa, nattino2020assessing}. The fact that a $p$-value dose not measure the size of an effect -- in our context, the degree of the  inconsistency between the working model and observed data -- undermines the usefulness of goodness-of-fit tests in general. 

In today's business world, the general interest has shifted from making a ``yes/no'' decision to knowing {\it why, how, and what to do}. As for model assessments, diagnostic tools are much needed to reveal reasons and inform actions with interpretable evidence. Our development in this paper responds to this need by (i) offering easy-to-interpret plots for measuring scientific/business significance of model misspecification; and (ii) showing how to derive business insights and actionable clues for model treatment and improvement. Our work is motivated by the following business case studies (see Section 4 for details).

\begin{Case}[Wine Marketing] To better serve wine consumers, there is a recent effort to use analytics to understand how physicochemical test result can determine humane tasting preference. Physicochemical tests are carried out in labs and measure $pH$, $alcohol$, and many other characteristics of wine, whereas humane tasting preference is subjective and rated by wine experts. Understanding how wine characteristics influence tasting preference can help inform and guide wine  manufacturing and marketing. As an ordinal scale (e.g., 0-10) is often used to measure human preference, ordinal regression models (e.g., adjacent-categorical model) can generate interpretable insights for wine manufacturers. But a general diagnostic procedure is needed to guide, assess, and refine the model building process. 
\end{Case}

\begin{Case}[Bike Sharing] Bike sharing is a revolution of the traditional bike rental business. It allows users to rent a bike from one of the multiple rental locations and return it back to another location close to his/her destination. To improve the efficiency of the rental system, it is crucial to examine how weather conditions and time/day influence consumer behavior. As the outcome is the number of hourly rentals during a day, Poisson regression models can generate interpretable insights for system managers. But again, a general diagnostic procedure is needed to guide, assess, and refine the model building process.
\end{Case}
	
In this paper, we develop a new notion of residuals for discrete data regression. Unlike the literature that always defines a single-valued quantity as the residual, we propose to {\it use a function as a vehicle to retain the residual information}. In the presence of data discreteness, we show that a functional residual is an appropriate tool for summarizing the residual randomness that cannot be captured by the structural part of the model. We establish theoretical properties of the functional residual for general discrete data regression models. These properties lead to the innovation of new diagnostic tools, such as the functional-residual-{\it vs}-covariate plot and Function-to-Function ({\it Fn-Fn}) plot. These tools play similar roles of traditional residual-{\it vs}-covariate plot and Quantile-Quantile (Q-Q) plot for assessing linear regression models. Our extensive simulation studies demonstrate that the use of our new diagnostic tools can reveal a variety of model misspecifications, such as not properly including a higher-order term, an explanatory variable, an interaction effect, a dispersion parameter, or a zero-inflation component. The effectiveness of our methodology is also demonstrated in the two case studies of wine marketing ($n=4898$) and bike sharing ($n=8734$). In each case study, we show that our tools generate useful statistical and business insights, which suggest model treatment and guide the entire model refining process. 

Our framework broadens the diagnostic scope as it applies to virtually all parametric models for binary, ordinal and count data, all in a unified diagnostic scheme. This strength distinguishes our methodology from the vast majority of model assessment methods that are either data-specific or model-specific. Our theoretical and graphical results can be interpreted in the same way irrespective to what type of data or model are dealt with. In fact, these results are similar to what has been established for linear regression models. This similarity bridges the interpretability gap between discrete-data and continuous-data regression models. 

In Section 5, we further show that our functional residual can serve as a vehicle to unify recently developed sign-based residual \citep{li2010test,li2012new} and surrogate residual \citep{liu2018residuals,liu2020assessing}. In fact, these residuals are simply point statistics that can be drawn from our functional residual.

\section{A unifying framework for model diagnostics}
\subsection{A general model setting for discrete data}
Let $Y$ be a discrete outcome and $\Xi$ the set of its values with a positive probability. Examples include 
\begin{itemize}
 \item a binary variable $Y$ with $\Xi=\{0,1\}$;
 \item an ordinal variable $Y$ with $\Xi=\{0,1,\ldots,J\}$; and 
 \item a count variable $Y$ with $\Xi=\{0,1,2,\ldots\}$.
\end{itemize}
Given a set of explanatory or predictive variables $\bm X=\{X_1,\ldots,X_p\}$, researchers and practitioners may use their domain knowledge and conventions to initiate a parametric model to quantitatively characterize the influence of $\bm X$ on the probabilistic distribution of $Y$. In its most general form, such a {\it working model} can be written as 
\begin{equation}
  \label{eq:general-model}
  Y \mid \bm X \sim \pi(y; \bm X, \bm\beta),
\end{equation} 
where $\pi(y; \bm X, \bm\beta)=\Pr\{Y \leq y \mid \bm X, \bm\beta\}$ is a discrete distribution function defined on the sample space $\Xi$. Model~\eqref{eq:general-model} encompasses commonly used models for discrete variables, such as logistic/probit regression models for binary data, ordered logit/probit regression models for ordinal data, and Poisson regression models for count data. We stress that Model~\eqref{eq:general-model} is broader than the class of generalized linear models, and it includes the adjacent-category logit model, stereotype logit model, and other non-linear models. 

Our goal is to assess whether or not the working Model \eqref{eq:general-model} is consistent with the observed data set $\{(y_i,\bm x_i):i=1,2,\ldots,n\}$. Assuming that the data are generated from the true distribution $\pi_0(y;\bm X)$, we are seeking statistical evidence against the working assumption of $\pi(\cdot; \bm X) \equiv \pi_0(\cdot; \bm X)$ in model diagnostics. If such evidence is present, either numerically or graphically, we hope that it can give us clues for  model treatment, such as adding or modifying certain components to adjust or refine the model.

\subsection{Functional residual: intuition and definition}
We introduce a new notion of residual for discrete data regression as in the form of Model~\eqref{eq:general-model}. Unlike classical residuals (Pearson or deviance residuals included) which merely use a {\it single point} statistic, we propose to use a {\it function} to represent residual information, which is termed as a {\it functional residual}. Although it is specifically proposed for discrete data regression models, the notion is consistent with the ordinary residual used for linear regression models. Our intuition originates from the two fundamental observations below. 

\noindent(Obs-1) {\it A residual is a quantity describing the residual randomness that cannot be captured by the structural part of the model. } 

\begin{figure}
\centering
\includegraphics[width =1.0\textwidth]{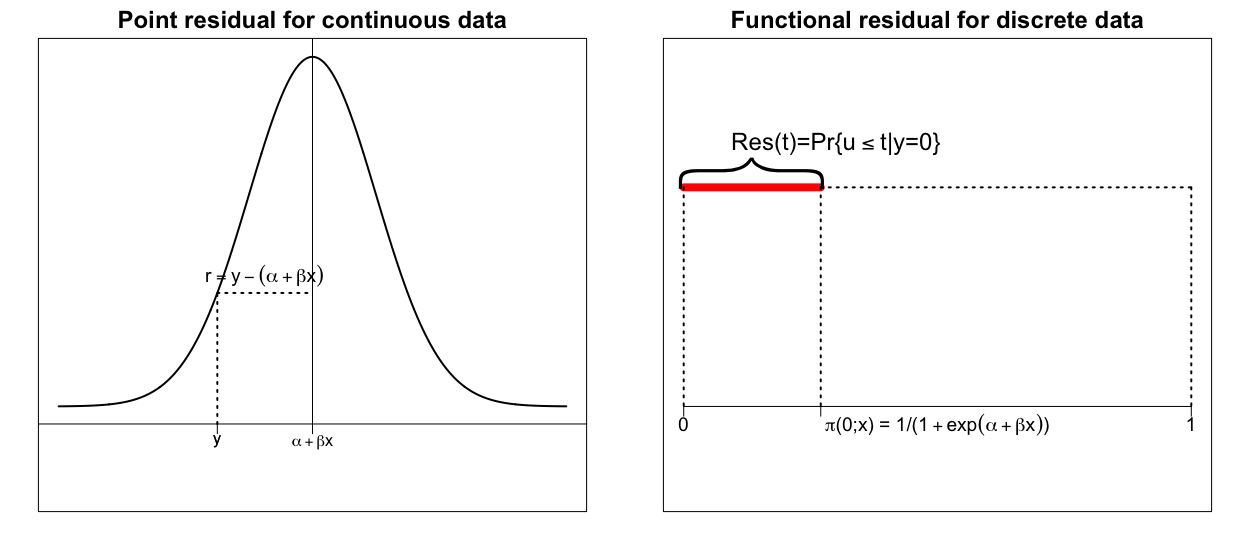}
\caption{An illustration of our idea of constructing a functional residual for a logistic regression model (right panel), in comparison to the traditional method of deriving a point residual for a linear regression model (left panel). The entire thickened red line segment represents the functional residual for the observation $(y=0, x=1)$.}
\label{fig:idea}
\end{figure}

\noindent This principle is followed in finding a residual for the linear regression model $Y=\alpha + \beta X + \epsilon$. The ordinary residual is $\{y-(\alpha + \beta x)\}$, the difference between the observed $y$ and the structural part $(\alpha + \beta x)$ of the model. This quantity reflects the residual randomness that cannot be captured by the structural part $(\alpha + \beta x)$ (See the left panel of Figure \ref{fig:idea}). For discrete data, the question is: how can we follow (Obs-1) to develop a general notion of residuals? The answer follows the second observation stated below.

\noindent(Obs-2) {\it Given $\bm X=\bm x$, the distribution $\pi(y;\bm x)$ specified by Model~\eqref{eq:general-model} gives nothing but a partition of the probability interval [0,1]. The structural part of this model is no more than the totality of the cutpoints $\{\pi(0;\bm x),\pi(1;\bm x),\ldots\}$.}

\noindent To elaborate on (Obs-2), we consider a binary outcome which follows a logistic regression model as below
\begin{equation}
\label{eq:logistic}
\text{logit}(\Pr\{Y=1\}) = -1 + 2 X.
\end{equation}

Given $X=x=1$, for instance, the logistic model essentially gives one probability $\pi(0;x) = \Pr\{Y=0 \mid x \} = 0.27$. As illustrated in the right panel of Figure~\ref{fig:idea}, this quantity yields a partition of the probability interval [0,1], with the cutpoint being $\pi(0;x) =0.27$. Figure~\ref{fig:idea} suggests that this quantity plays a role similar to $(\alpha + \beta x)$ in the linear regression model, and it is the structural part of the logistic regression. A question follows: given an observed outcome, say, $y=0$, what is the residual randomness that {\it cannot} be captured by $\pi(0;x)$, and how to describe this residual randomness?

To answer this question, we let $U$ be a random variable whose value is randomly scattered onto the interval [0,1]. The binary outcome $Y$, whose probabilistic behavior is described by the logistic model, can be equivalently expressed as 
\begin{eqnarray*}
Y=0 \mid X=x\ \text{if and only if}\ 0 < U \leq \pi(0;x);\ \text{and}\\ 
Y=1 \mid X=x\ \text{if and only if}\  \pi(0;x) < U \leq 1.
\end{eqnarray*}
In other words, given an observed outcome $y=0$, what we know is that the realization of $U$ falls in the interval $(0, \pi(0;x))$, yet what we do not know is where it is actually located. Following this line of thought, the residual randomness can be characterized by a truncated distribution $U \mid 0 < U \leq \pi(0;x)$, which is a uniform distribution $U(0, \pi(0;x))$. In this situation, obviously, a single-valued quantity, such as a traditional residual statistic, is not adequate to retain all the residual information. A {\it functional quantity} is more appropriate as an information carrier.

Specifically, given an observation $(y,x)$ of the logistic regression model \eqref{eq:logistic}, we define its {\it functional residual} as 
$$ Res (t; y,x) = \left\{
  \begin{array}{l l}
    F_{U(0, \pi(0;x))}(t)=\Pr\{ U(0, \pi(0;x)) \leq t\} & \quad \text{if $y = 0$}, \\
    F_{U(\pi(0;x),1)}(t)=\Pr\{ U(\pi(0;x),1) \leq t\} & \quad \text{if $y = 1$}.
  \end{array} \right.$$
Apparently, a functional residual is a mapping from the sample space $\Omega$ to a function space consisting of all the cumulative distribution functions (CDFs). As a concrete example, the functional residual for $(y=0,x=1)$ is 
$$ Res (t; y=0,x=1) = \Pr\{ U(0, 0.27) \leq t\} = \left\{
  \begin{array}{l l}
    \frac{t}{0.27} & \quad \text{if $0 < t \leq 0.27$}, \\
    1 & \quad \text{if $0.27 < t \leq 1$}.
  \end{array} \right.$$
The functional residual for $(y=1,x=-1)$ is 
$$ Res (t; y=1,x=-1) = \Pr\{ U(0.95, 1) \leq t\} = \left\{
  \begin{array}{l l}
    0 & \quad \text{if $0 < t \leq 0.95$}, \\
    \frac{t-0.95}{1-0.95} & \quad \text{if $0.95 < t \leq 1$}.
  \end{array} \right.$$
  
\begin{figure}[]
\centering
\includegraphics[width =1.0\textwidth]{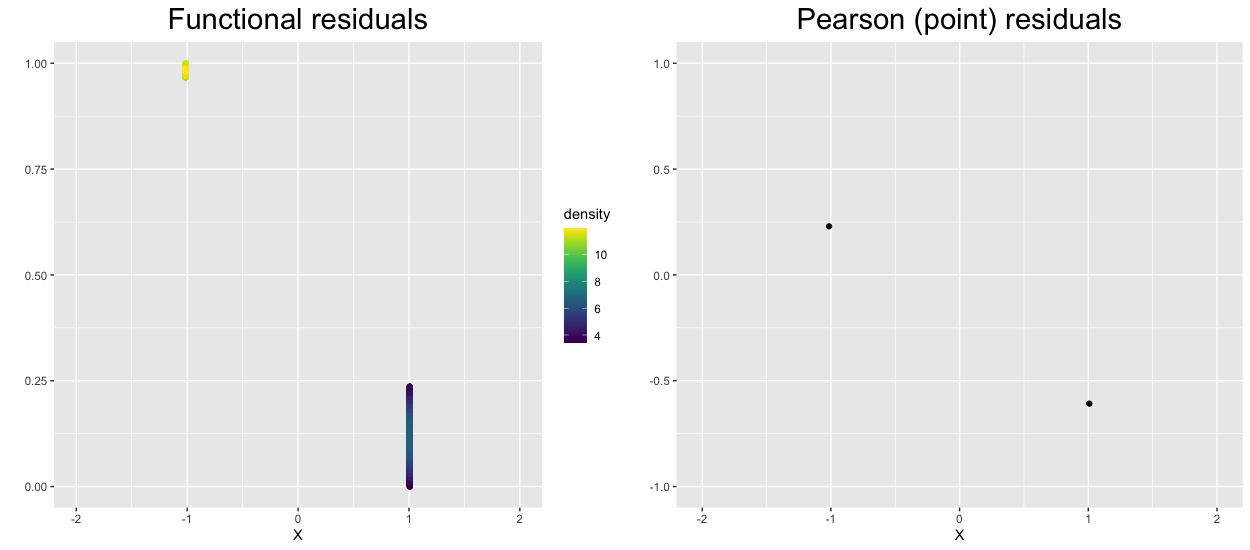}
\caption{A comparison of functional residuals and Pearson (point) residuals when the observation is $(y=0,x=1)$ or $(y=1,x=-1)$.}
\label{fig:illustration}
\end{figure}

The left panel of Figure~\ref{fig:illustration} illustrates these two functional residuals in their density forms. It is manifest that the use of the whole function is distinct from the traditional use of a single point as residual statistic, such as Pearson residual in the right panel of Figure \ref{fig:illustration}. Below, we present a general notion of functional residuals for discrete data. 

\begin{definition}[Functional Residual]
For a discrete outcome $Y$ believed to follow Model~\eqref{eq:general-model} with a set of explanatory variables $\bm X$, a functional residual for an observation $(y, \bm x)$ is a mapping from the sample space $\Omega$ to the function space $\Pi = \{F(t): 0 \leq t \leq 1; 0\leq F(t) \leq 1;\  \text{and}\ F(t_1)\leq F(t_2)\ \text{for any}\ t_1 < t_2\}$. Specifically, 
\begin{equation}
\label{eq:res-func}
(y, \bm x) \to  Res (t; y,\bm x) = F_{U(\pi(y-1;\bm x), \pi(y;\bm x))}(t)=\Pr\{ U(\pi(y-1;\bm x), \pi(y;\bm x)) \leq t \}.
\end{equation}
\end{definition}

Intuitively, a functional residual is loaded with more information than a single-valued residual statistic.  In the rest of this section, we will establish its theoretical properties and develop new function-based diagnostic tools. 

\subsection{Theoretical properties and new diagnostic tools}
In this subsection, we examine the conditional and unconditional expectation of the functional residual defined in \eqref{eq:res-func}. The theoretical results lead to the development of function-based diagnostic plots, which will be examined in simulated examples and real case studies. 

\begin{theorem}[Conditional Expectation under the Null]
Given $\bm X=\bm x$, the conditional expectation of the functional residual $Res(\cdot; Y, \bm x)$ in \eqref{eq:res-func} is the CDF of a U(0,1) distribution, i.e.,
\[
E_Y Res(t; Y, \bm x) = F_{U(0,1)}(t)=t\ \text{for any}\ t \in [0,1], 
\]
provided that $\pi(\cdot; \bm x) \equiv \pi_0(\cdot; \bm x)$.
\label{thm:conditional-exp}
\end{theorem}

On the basis of Theorem~\ref{thm:conditional-exp}, we propose to use a functional-residual-{\it vs}-covariate plot to examine the working model. Unlike the traditional residual-{\it vs}-covariate plot that displays point  residual statistics, a  functional-residual-{\it vs}-covariate plot exhibits the density function $\partial Res(t;y,x)/\partial t$ versus $x$. Such a new tool can be better visualized through the lens of a heat map, which is demonstrated in the following example.  

\begin{example}[Working model specified correctly]
\label{Exp:ordinal-correct}
We simulate 1000 ordinal data points $y_i\ (=0,1,2,3,4)$ from an  adjacent-category logit model 
$$log\frac{\Pr\{Y=j\}}{\Pr\{Y=j+1\}}=\alpha_j+\beta_1 X+\beta_2 X^2,  \qquad  j=0,1,2,3$$ 
where $(\alpha_0, \alpha_1, \alpha_2, \alpha_3)=(1.5, 1.5, -1, 1)$, $(\beta_1,\beta_2)=(1.5,-1)$ and the covariate $X \sim \mathcal{N}(0,1)$. Suppose the working model is specified correctly. We obtain functional residuals as defined in \eqref{eq:res-func}, and cast its density form $\partial Res(t;y_i,x_i)/\partial t$ against $x_i$. This functional-residual-{\it vs}-covariate plot is rendered in a heat map as shown in the Figure~\ref{fig:ordinal-correct}(a). It indicates that the ``heat'' is evenly distributed on the two sides of the center horizontal line. This observation confirms the result in Theorem~\ref{thm:conditional-exp}; that is, the null distribution of the ``heat'' is uniformly distributed on the (0,1) interval. The functional-residual-{\it vs}-covariate plot can also be rendered on the scale of a normal distribution as seen in the Figure \ref{fig:ordinal-correct}(b), which exhibits a similar pattern of a symmetric distribution of the ``heat''. In contrast, when the classical deviance and Pearson residuals are used, the traditional residual-{\it vs}-covariate plots exhibit {\it asymmetric} distributions around the center horizontal line (see Figure~\ref{fig:ordinal-correct}(c) and (d)). This observation once again confirms a phenomenon discussed in \cite{liu2018residuals} and \cite{liu2020assessing}; that is, {\it classical residuals may exhibit unusual patterns even when the model is specified correctly}. The same problem is seen in other point residuals, such as the sign-based residual \citep{li2012new} and the generalized residual \citep{franses2001quantitative}.

\begin{figure}
\centering
\includegraphics[width =1.0\textwidth]{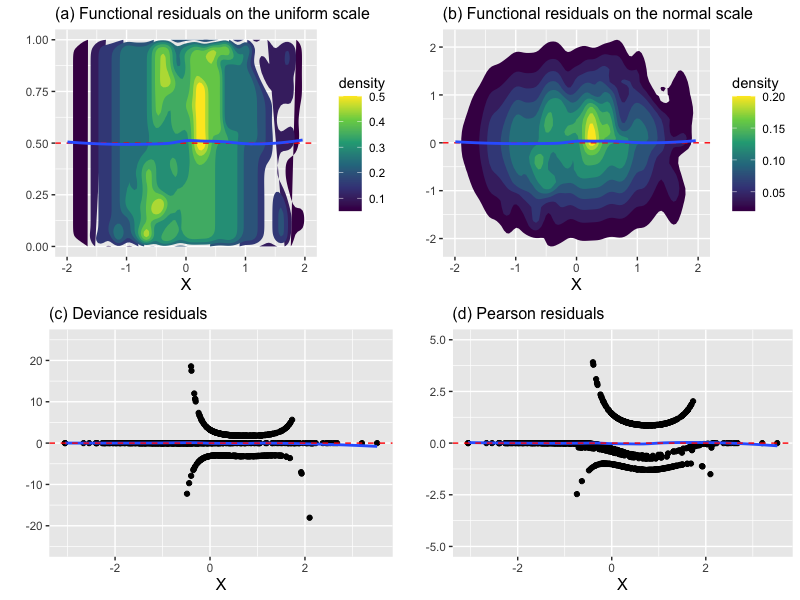}
\caption{Proposed functional-residual-{\it vs}-covariate plots (upper low) and traditional residual-{\it vs}-covariate plots when the working model is specified correctly for ordinal data in the setting of Example~\ref{Exp:ordinal-correct}.}
\label{fig:ordinal-correct}
\end{figure}

\end{example}

In what follows, we examine the unconditional expectation of the functional residual $Res(\cdot; Y, \bm X)$. The result can be used to develop a new diagnostic tool that plays a similar role of Quantile-Quantile (Q-Q) plots in traditional regression analysis.
\begin{theorem}[Unconditional Expectation under the Null]
The unconditional expectation of the functional residual $Res(\cdot; Y, \bm X)$ in \eqref{eq:res-func} is the CDF of a U(0,1) distribution, i.e.,
\[
E_{(Y,\bm X)} Res(t; Y, \bm X) = F_{U(0,1)}(t)=t\ \text{for any}\ t \in [0,1],
\]
provided that $\pi(\cdot; \bm X=\bm x) \equiv \pi_0(\cdot; \bm X=\bm x)$ for any $\bm x$.
\label{thm:unconditional-exp}
\end{theorem}

Following the law of large numbers, we have the following corollary immediately. 

\begin{corollary}
 Suppose $(Y_1,\bm X_1), (Y_2,\bm X_2), \ldots$ is an infinite sequence of i.i.d. random variables. Then, for any $t \in (0,1)$, 
\begin{equation}
\overline{Res}(t)=\frac{1}{n}\sum_{i=1}^n Res(t;Y_i,\bm X_i) \to F_{U(0,1)}(t)=t \quad \text{almost surely},
\end{equation}
provided that $\pi(\cdot; \bm X=\bm x) \equiv \pi_0(\cdot; \bm X=\bm x)$ for any $\bm x$.
\label{cor:unconditional-ave}
\end{corollary}

On the basis of Corollary~\ref{cor:unconditional-ave}, we propose to draw the function $\overline{Res}(t)$ against its null function $F_{U(0,1)}(t)=t$, which results in a {\it Function-Function (Fn-Fn) plot}. If the working model is specified (approximately) correctly, the {\it Fn-Fn} plot should yield a curve aligning (approximately) with the 45-degree straight line. With this regard, the {\it Fn-Fn} plot can be used in analogy to the classical Q-Q plot for linear regression models. As there is no well-established and widely-accepted Q-Q plots for discrete data models, our {\it Fn-Fn} plot fills the void while maintaining the similar utility. 

\setcounter{example}{0}
\begin{example}[Continued]
We draw {\it Fn-Fn} plots for the same adjacent-category logit model considered previously. When the working model is specified correctly, Figure~\ref{fig:Fn-Fn-correct}(a) shows the {\it Fn-Fn} curve aligns almost perfectly with the 45-degree line. This observation confirms the result in Corollary~\ref{cor:unconditional-ave}. As a futher investigation, we restrict the sample to a subset $\{(y_i,x_i) \mid x_i < 0\}$ and reproduce the {\it Fn-Fn} plot. As seen in Figure~\ref{fig:Fn-Fn-correct}(b), the {\it Fn-Fn} curve still aligns with the 45-degree line very well. The probe of specific subgroups as such may be useful in practice. 

\begin{figure}[]
\centering
\includegraphics[width =1.0\textwidth]{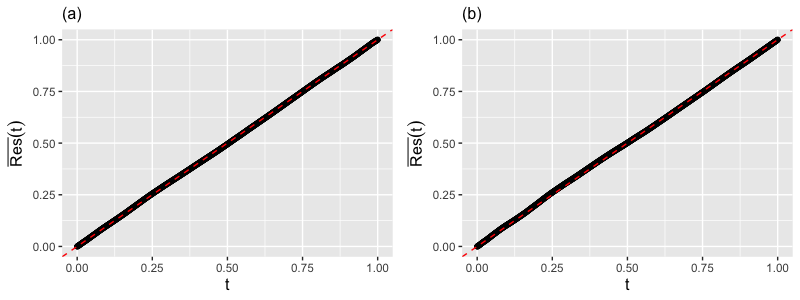}
\caption{Proposed {\it Fn-Fn} plots when the working model is specified correctly for ordinal data in the setting of Example~\ref{Exp:ordinal-correct}. The left and right panels are obtained using the full sample (1000 observations) and the subsample $\{(y_i,x_i) \mid x_i < 0\}$, respectively. }
\label{fig:Fn-Fn-correct}
\end{figure}

\end{example}

\begin{theorem}[Uniform convergence]
Suppose $(Y_1,\bm X_1), (Y_2,\bm X_2), \ldots$ is an infinite sequence of i.i.d. random variables. The average of the functional residuals $Res(t; Y_i, \bm X_i)$ in \eqref{eq:res-func} converges to the function $F_{U(0,1)}(t)=t$ uniformly; that is, for any $\epsilon > 0$,
$$\sup_{t \in [0,1]}P(|\overline{Res}(t;Y_n,\bm X_n)-t| < \epsilon) \xrightarrow{} 1 , \quad\text{as } n \xrightarrow{} \infty .$$

\label{the:wlln}
\end{theorem}

As compared to Theorem \ref{thm:unconditional-exp} and Corollary \ref{cor:unconditional-ave}, which only justify the point-to-point convergence of $\overline{Res}(t;Y_n,\bm X_n)$ to $t$, Theorem \ref{the:wlln} further shows the convergence is uniform for the entire set $t \in [0,1]$. The implication is when the model is right, the entire curve $(t,\overline{Res}(t))$ will move toward the 45-degree line at the same rate as the sample size increases. In other words, if the curve $(t,\overline{Res}(t))$ deviates partly from the 45-degree line, it is not a result of different convergence rates over $t \in [0,1]$ but an indication of model misspecification.

The development of Theorem 1-3 does not depend on the type of outcome data or the form of working model. This allows us to use our diagnostic tools  in a similar way to the common practice of using diagnostic plots for linear regression models.

\section{Detecting model misspecifications}
In this section, we use simulated examples to demonstrate the utility of our diagnostic tools in the detection of a variety of model misspecifications. The examples will manifest two features of the functional-residual-based diagnostic framework:
\begin{itemize}
\item Unlike hypothesis testing that merely yields a ``yes/no" answer, our approach reveals the {\it extent} to which the working model deviates from the true model, and furthermore, it gives {\it actionable clues} as to what model component is misspecified as well as what treatment should be applied to improve the model.
\item Unlike hypothesis testing procedures that are data- or model-specific, our approach provides a unifying treatment in the sense that  diagnostic results can be interpreted in a similar fashion {\it regardless of the type of response data (binary, ordinal, or count) and the choice of regression models}. 
\end{itemize}

\subsection{Regression models for ordinal data}
Different from \cite{liu2018residuals} which focused on the cumulative link model, our approach is not constrained to latent-variable-induced models. Instead, it applies to any model for ordinal data. In this subsection, we use the adjacent-category logit model to study an ordinal outcome ($Y=0,1,2,3,4$) consistently throughout the examples.

\begin{figure}[h]
\centering
\includegraphics[width =1\textwidth]{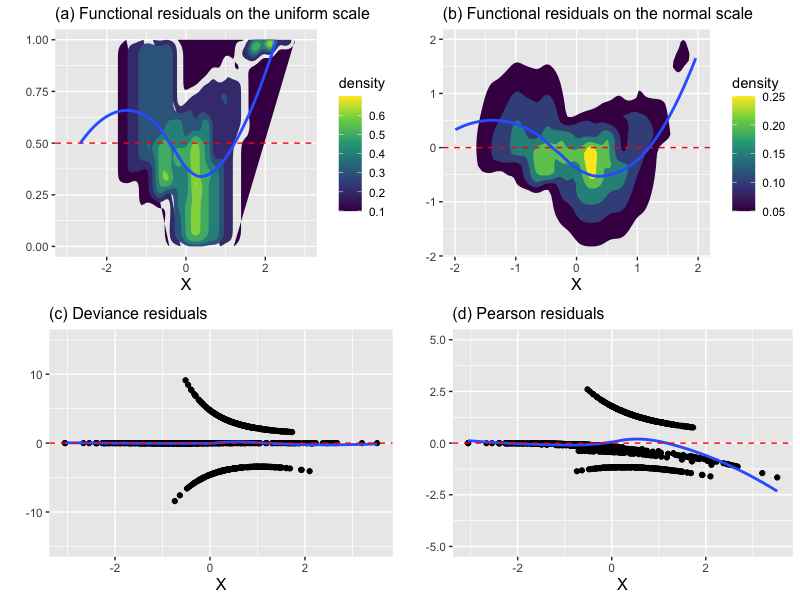}
\caption{Proposed functional-residual-{\it vs}-covariate plots (upper row) and traditional residual-{\it vs}-covariate plots (lower row) when the quadratic term $X^2$ is missing in the working adjacent-category logit model (Example~\ref{Exp:ordinal-quadratic}).}
\label{fig:ordinal-quadratic}
\end{figure}

\begin{figure}[]
\centering
\includegraphics[width =0.6\textwidth]{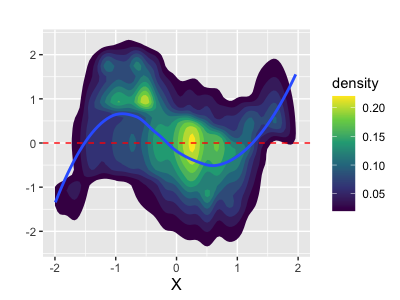}
\caption{Functional-residual-{\it vs}-covariate plot when the cubic term $X^3$ is missing in the working adjacent-category logit model (Example~\ref{Exp:ordinal-quadratic}).}
\label{fig:ordinal-cubic}
\end{figure}

\begin{example}[Missing of a higher order term]
\label{Exp:ordinal-quadratic}
The ordinal data are generated from the same adjacent-category logit model as in Example~\ref{Exp:ordinal-correct}, where the $X^2$ term is a crucial component in the underlying model. Initial analysis, however, often begins with a simple model that only contain a linear term of $X$, such as 
$log(p_j/p_{j+1})=\alpha_j+\beta  X$, where $p_j=\Pr\{Y=j\}$.
For this working model, we obtain its functional-residual-{\it vs}-covariate plots, which are displayed in Figure~\ref{fig:ordinal-quadratic}(a). The ``heat'' in the upper left panel is unevenly distributed around the center line, being quite different from that in Figure~\ref{fig:ordinal-correct}(a) where the model is specified correctly. Moreover, the LOWESS curve clearly indicates the missing of the quadratic term in the working model. Visualized in Figure~\ref{fig:ordinal-quadratic}(b) are functional residuals {\it in the standard normal scale} (see Remark \ref{remark_1}), which exhibits a similar pattern. For comparison purposes, classical residual-{\it vs}-covariate plots are included in the lower row of Figure~\ref{fig:ordinal-quadratic}. The deviance residuals in Figure~\ref{fig:ordinal-quadratic}(c) barely yield any evidence against the use of the working model without the quadratic term. In fact, the comparison of the plot here with that in Figure~\ref{fig:ordinal-correct}(c) manifests a ``fatal defect'' of deviance residuals, as we can hardly tell which residual-{\it vs}-covariate plot is normal and which is abnormal. In Figure~\ref{fig:ordinal-quadratic}(d), Pearson residuals exhibit a declining trend in the region where $X > 0$. But it may be confusing if we compare this plot with Figure~\ref{fig:ordinal-correct}(d) as "U"-shaped curves are also seen in the latter.

As a further study, we simulate 1000 samples from a cubit model $log(p_j/p_{j+1})=\alpha_j+\beta_1  X+ \beta_2 X^2+\beta_3 X^3$, where $(\alpha_0, \alpha_1, \alpha_2, \alpha_3)=(-1, 1.5, 2, 3)$, $(\beta_1,\beta_2,\beta_3)=(2,-1,-1.5)$, and $X \sim \mathcal{N}(0,1)$. Assuming that the quadratic term is included in the working model while the cubit term is not, we cast its functional residuals to the normal scale in Figure~\ref{fig:ordinal-cubic}. The cubic LOWESS curve there informs that a quadratic polynomial is not adequate for explaining $log(p_j/p_{j+1})$ and a high-order term should be added.

\begin{remark}
\label{remark_1}
 Although the functional-residual-{\it vs}-covariate plots on the uniform and normal scales are loaded with the same amount of  information, we recommend to use the normal scale for better visualization purposes. We may hardly distinguish points, such as 0.99 and 0.999, on the edges of the (0,1) interval, but the difference is apparent on the normal scale as $\Phi^{-1}(0.99)=2.33$ and $\Phi^{-1}(0.999)=3.09$. We will use the normal scale to present functional residuals in the rest of our paper.
 
\end{remark}

\end{example}

\begin{figure}[]
\centering
\includegraphics[width =1.0\textwidth]{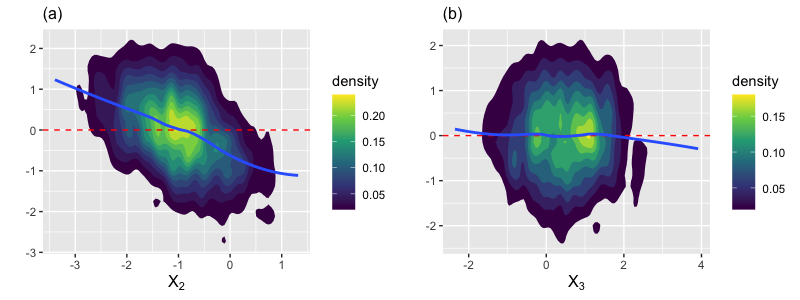}
\caption{Functional-residual-{\it vs}-covariate plots when $X_2$ is correlated with ordinal data $Y$ (the left panel) whereas $X_3$ is not (the right panel) in the setting of Example~\ref{Exp:ordinal-covariate}.}
\label{fig:ordinal-covariate}
\end{figure}

\begin{example}[Missing of relevant covariates]
\label{Exp:ordinal-covariate}
We simulate 1000 samples from the model $log(p_j/p_{j+1})=\alpha_j+\beta_1  X_1+ \beta_2 X_2 + \beta_3 X_3$. Here, $(\beta_1, \beta_2, \beta_3)=(1.5, 1, 0)$, $(\alpha_0, \alpha_1, \alpha_2, \alpha_3)=(-1, -2, 0.5, 2)$, $X_1 \sim  \mathcal{N}(0,1)$, $X_2 \sim  \mathcal{N}(-1,0.8)$, and $X_3 \sim \mathcal{N}(0.5, 1)$. In this setting, both $X_1$ and $X_2$ are correlated with the ordinal response $Y$, whereas $X_3$ is irrelevant as $\beta_3=0$. Assuming that the working model only contains the covariate $X_1$, we obtain the functional residuals. Figure~\ref{fig:ordinal-covariate} casts them against $X_2$ and $X_3$ that are not included in the working model. The strong correlation seen in Figure~\ref{fig:ordinal-covariate}(a) indicates that the missing covariate $X_2$ may help explain a significant proportion of the residual variability, and thus it should be added to the model. To the contrary, the symmetrically distributed ``heat'' in Figure~\ref{fig:ordinal-covariate}(b) does not exhibit any correlation between our functional residual and $X_3$, which is consistent with the fact that $X_3$ is irrelevant ($\beta_3=0$).

\end{example}

\begin{figure}[]
\centering
\includegraphics[width =1.0\textwidth]{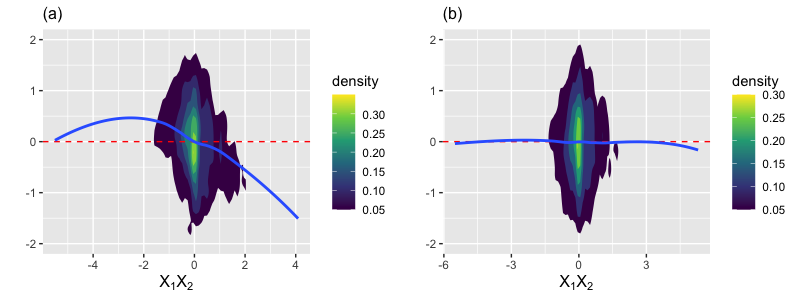}
\caption{Functional-residual-{\it vs}-covariate plots before (left) and after (right) the interaction term $X_1 X_2$ is included in the working adjacent-category logit model (Example~\ref{Exp:ordinal-interaction}).}
\label{fig:ordinal-interaction}
\end{figure}

\begin{example}[Missing of interaction]
\label{Exp:ordinal-interaction}
We simulate 1000 samples from the model 
$log(p_j/p_{j+1})=\alpha_j+\beta_1  X_1+ \beta_2 X_2+\beta_3 X_1 X_2$, which contains an interaction effect between the two covariates. Here, $(\beta_1, \beta_2, \beta_3)=(1, 2, 2)$, $(\alpha_0, \alpha_1, \alpha_2, \alpha_3)=(-1, -2, 0.5, 2)$, $X_1 \sim  \mathcal{N}(0,1)$, and  $X_2 \sim  \mathcal{N}(-1,0.8)$. Assume that the interaction is not built into the initial model. We obtain functional residuals and cast them against the product of $X_1$ and $X_2$ in Figure~\ref{fig:ordinal-interaction}(a). The decreasing trend is clear evidence that the residual variability can be partially explained by the missing interaction term. After adding $X_1X_2$ to the working model, the updated functional-residual-{\it vs}-covariate plot in Figure~\ref{fig:ordinal-interaction}(b) shows that the decreasing trend disappears.

\end{example}

\subsection{Regression models for count data}
\begin{figure}[]
\centering
\includegraphics[width =1\textwidth]{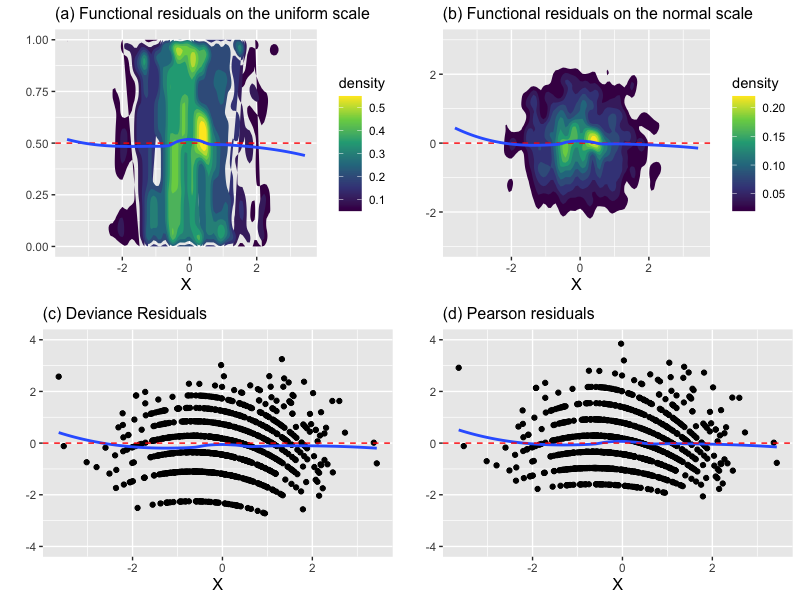}
\caption{Plots of the proposed functional residual (upper row) and traditional deviance and Pearson residuals (lower row) against the covariate $X$ when the working Poisson model is specified correctly for count data in the setting of Example~\ref{exp:poi1}.}
\label{fig:correctpoi}
\end{figure}

\begin{figure}[]
\centering
\includegraphics[width =1\textwidth]{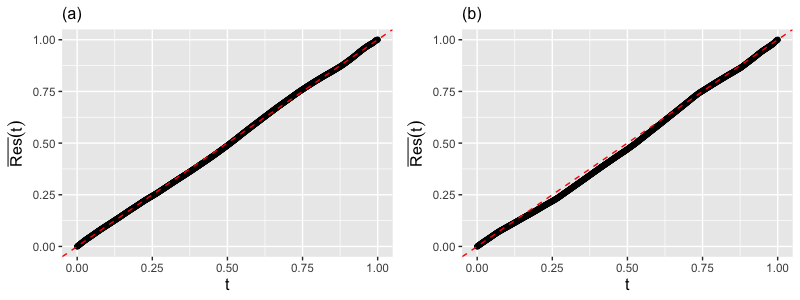}
\caption{Proposed {\it Fn-Fn} plots when the working model is specified correctly for count data in the setting of Example~\ref{exp:poi1}. The left and right panels are obtained using the full sample (1000 observations) and the subsample $\{(y_i,x_i) \mid x_i < 0\}$, respectively. }
\label{fig:fnfnpois}
\end{figure}

In this subsection, we continue to study the diagnostic utility of our functional residuals yet for integer-valued count data. Considered in the numerical examples are a variety of Poisson regression models. 
\begin{example}[Correct specification of the model]
\label{exp:poi1}
We generate 1000 samples from the Poisson regression model with a quadratic term $log(E(Y|X))=\alpha+ \beta_1 X+\beta_2 X^2$, where $(\alpha, \beta_1, \beta_2)=(1, 0.2, 0.15)$ and the covariate $X \sim \mathcal{N}(0, 1)$. 
When the model is specified correctly, we obtain our functional residuals as well as traditional deviance and Pearson residuals. Throwing these residuals against the covariate $X$ yields diagnostic plots in Figure \ref{fig:correctpoi}. The upper row of Figure  \ref{fig:correctpoi} shows that our functional residuals are evenly distributed around the dashed horizontal line, regardless of the scales (being uniform or normal) used for visualization. Our $F_n$-$F_n$ plot in Figure~\ref{fig:fnfnpois} reveals nothing worth attention either. This no-show of unusual patters is what we anticipate to see as the model is specified correctly. In contrast, the deviance residuals in Figure \ref{fig:correctpoi}(c) (and Pearson residuals in Figure \ref{fig:correctpoi}(d)) cluster together and form multiple parabolas opening downwards. This pattern may mislead analysts to believe that certain high-order terms of $X$ need to be added to the model. This observation once again reinforces our finding in Example~\ref{Exp:ordinal-correct}; that is, the discreteness of data can cause deceptive clusters and shapes if traditional point residuals are used to produce diagnostic plots.

\end{example}

In Appendix A, we present our diagnostic results when a higher order term, a relevant covariate, or an interaction is missing in the working Poisson model for count data. These are the cases of misspecification studied in the previous subsection for ordinal data. Our finding is that although the data type and parametric model are different, our functional residual yields similar diagnostic plots, which thus can be interpreted in the same fashion. In what follows, we focus on two misspecification cases that are common in modeling count data.

\begin{figure}[]
\centering
\includegraphics[width =1\textwidth]{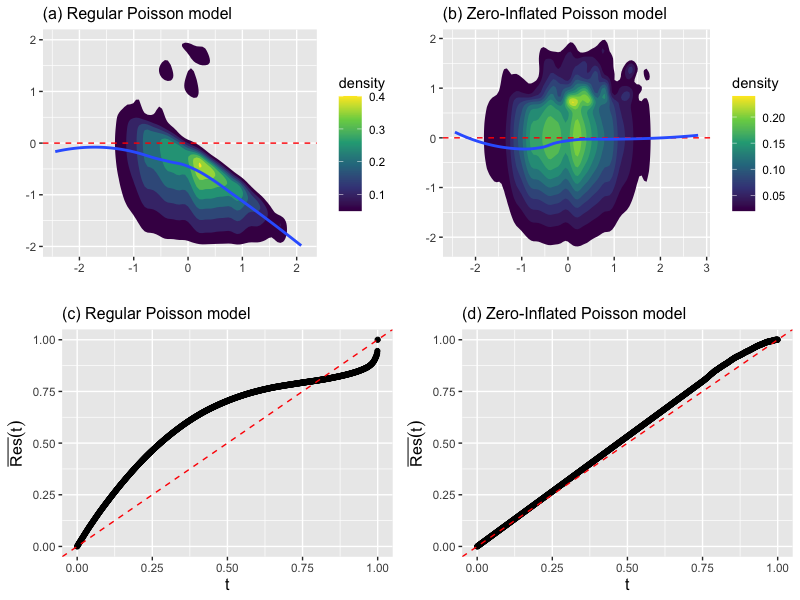}
\caption{Functional-residual-\textit{vs}-covariate plots (upper row) and \textit{Fn-Fn} plots (lower row) in the presence of excessive zeros. Shown are the results for the regular Poisson model (left) and zero-inflated Poisson model (right) used in Example \ref{zipmodel}.}
\label{fig:zip}
\end{figure}

\begin{example}[Missing of a zero-inflated component]
\label{zipmodel}
In the case of an excess of zeros in count data, the regular Poisson model is inadequate and a new component for modeling zeros needs to be introduced. For example, the count of zeros follows a logistic model $logit \left\{P(Y=0)\right\}=1+0.2 X$, while the non-zero counts follow a truncated Poisson model $log(E(Y|X))=1+X$ (for $Y \geq 1 $). We simulate 1000 samples from this so-called zero-inflated Poisson model with $X \sim  \mathcal{N}(0,0.8)$. We fit, however, a regular Poisson model to the simulated data. The functional-residual-\textit{vs}-covariate plot is displayed in Figure \ref{fig:zip}(a). It is observed that the vast majority of the functional residuals falls below the horizontal line at zero, and a long slice of the residual cloud approaches very close to $-2$. This is an indication that the proportions of smaller count values, such as zeros, are underestimated by the regular Poisson model. This observation is confirmed in Figure \ref{fig:zip}(c), where the lower tail of the \textit{Fn-Fn} curve goes far above the 45-degree line. We therefore update the model with an addition of a zero-inflated component and reproduce the functional-residual-\textit{vs}-covariate plot in Figure \ref{fig:zip}(b). The residual cloud becomes symmetrically distributed around zero and the LOWESS curve is almost flat. Figure \ref{fig:zip}(d) confirms the appropriateness of the model by showing that the \textit{Fn-Fn} curve aligns well with the 45-degree line.

\end{example}

\begin{figure}[]
\centering
\includegraphics[width =1\textwidth]{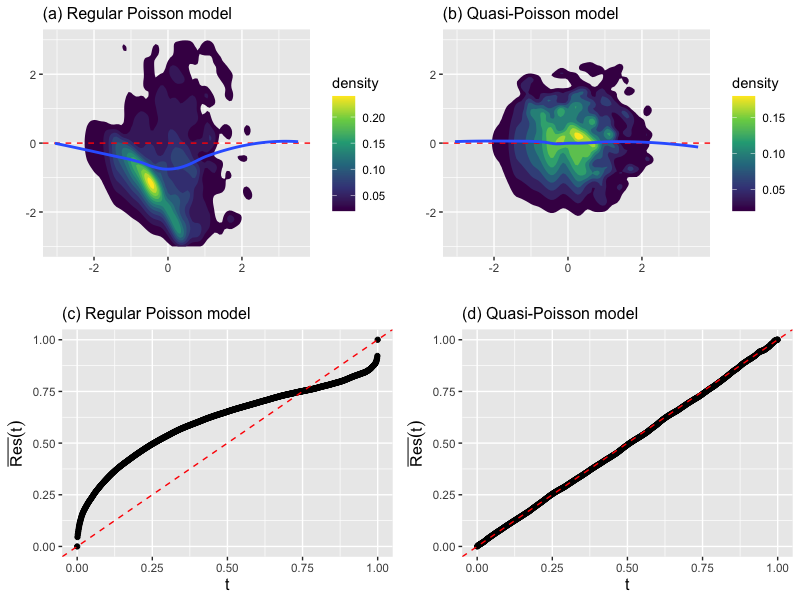}
\caption{Functional-residual-\textit{vs}-covariate plots (upper row) and \textit{Fn-Fn} plots (lower row)  in the presence of over-dispersion.  Shown are the results for regular Poisson model (left) and Quasi-Poisson model (right) used in Example \ref{overmodel}.}
\label{fig:overdis}
\end{figure}

\begin{example}[Missing of modeling dispersion]
\label{overmodel}

We simulate 1000 count data samples from a Poisson distribution with the dispersion parameter being $\frac{1+\phi}{\phi}=7$ where $log(E(Y|X))=1.2+1.3 X$ and $ X \sim  \mathcal{N}(0,1)$. The simulated data are fitted using a regular Poisson model with the default setting that the dispersion parameter is set as 1. In this case, the functional-residual-\textit{vs}-covariate plot in Figure \ref{fig:overdis}(a) shows that the residual cloud is stretched toward both the lower and higher ends. This is in contrast with the well-rounded residual cloud as seen in Figure \ref{fig:overdis}(b), where the over-dispersion is correctly modeled using a quasi-Poisson model. We also note that the over-dispersion in count data may result in inflated proportions of smaller count values (e.g., 0, 1, 2). This explains why in Figure \ref{fig:overdis}(c) the \textit{Fn-Fn} curve is highly skewed to the left. This skewness disappears in Figure \ref{fig:overdis}(d) when the dispersion parameter is introduced to capture the additional variability.

\end{example}

\section{Case studies}
In this section, we demonstrate the utility of functional residuals in statistical modeling for real business questions. Presented in our first case study is statistical modeling and diagnostics for wine quality to facilitate wine marketing and manufacturing, where the outcome variable expert rating is ordinal. The second case study also addresses a marketing problem, but the focus is on bike sharing and the outcome is the number of hourly rentals. Our analyses show that our diagnostic tools can (i) detect a variety of model misspecifications, which are otherwise concealed due to the discreteness of data; (ii) provide actionable clues that can guide model treatment; and (iii) be interpreted in the same way regardless of the type of outcomes or models.



\subsection{Diagnostics for wine quality modeling}

In order to better serve customers around the globe, the wine industry is investing and developing new technologies used in its manufacturing process. Among many key elements that can be analytically studied in this process are wine certification and quality assessment \citep{winesource}. The wine certification is performed in labs by using physicochemical tests which measure \textit{pH}, \textit{alcohol}, \textit{density}, and other characteristics of wine. However, tests of wine quality heavily rely on human experts, which is time-consuming and expensive. It is therefore desirable to establish the correlation between human preference and physicochemical properties. Understanding of this relationship can automate the assessment of wine quality and inform the entire wine manufacturing process. To this end, \cite{winesource} collected data of white vinho verde wine from the north of Portugal (\url{http://archive.ics.uci.edu/ml/datasets/Wine+Quality}). This data set consists of 4898 observations and 12 variables. The outcome {\it quality} is measured on a rating scale from 0 to 10, with 10 indicating the best quality. The observed ratings, ranging from 3 to 9, are derived from the judgment of wine experts. To understand how human ratings are influenced by the physicochemical characteristics of wine, we fit an adjacent-categorical model using 8 predictors listed in Table \ref{table1wine}.

\begin{figure}[]
\centering
\includegraphics[width =1\textwidth]{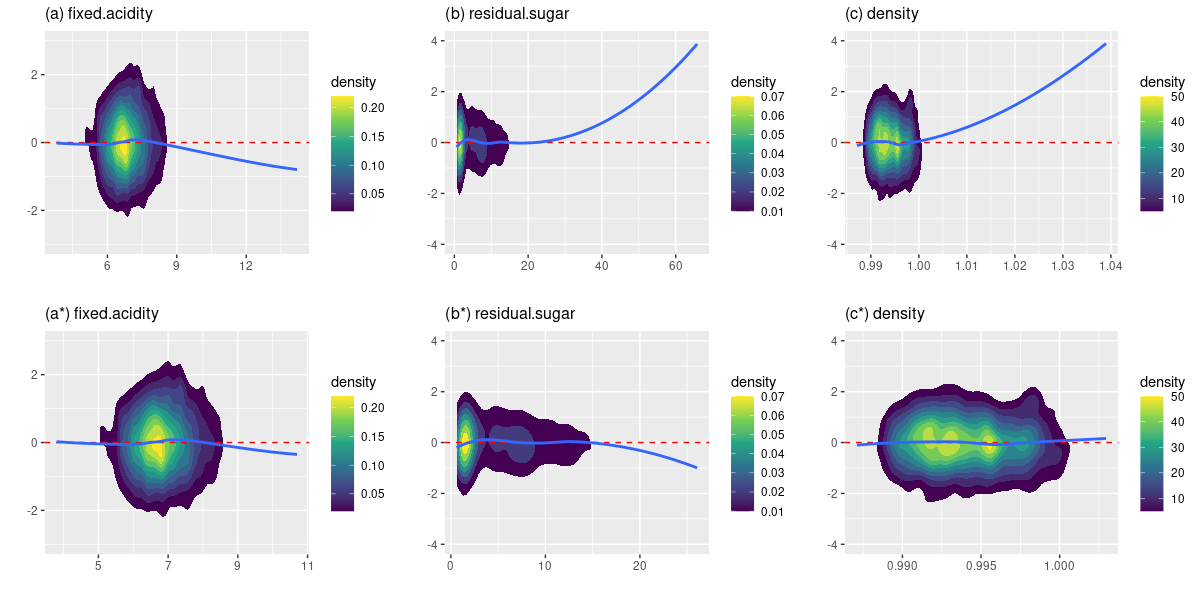}
\caption{Functional-residual-\textit{vs}-covariate plots before (upper row) and after (lower row) deleting the outliers from the wine quality dataset.}
\label{fig:firstheatmap}
\end{figure}

\begin{figure}[]
\centering
\includegraphics[width =1\textwidth]{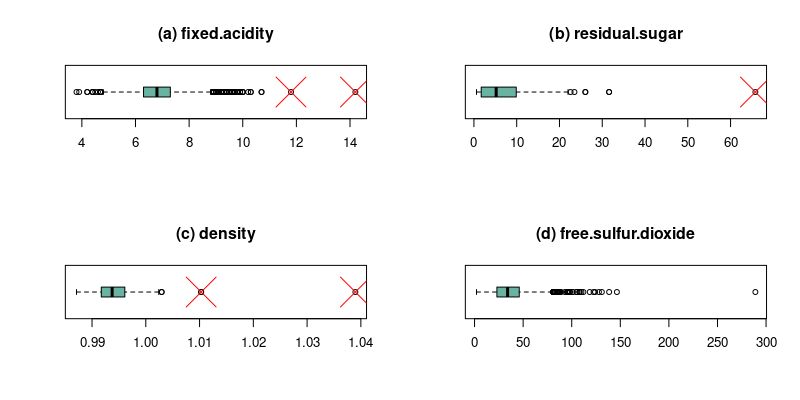}
\caption{Boxplots of the variables that may contain outliers in the wine quality dataset.}
\label{fig:boxplot}
\end{figure}

\begin{figure}[]
\centering
\includegraphics[width =1\textwidth]{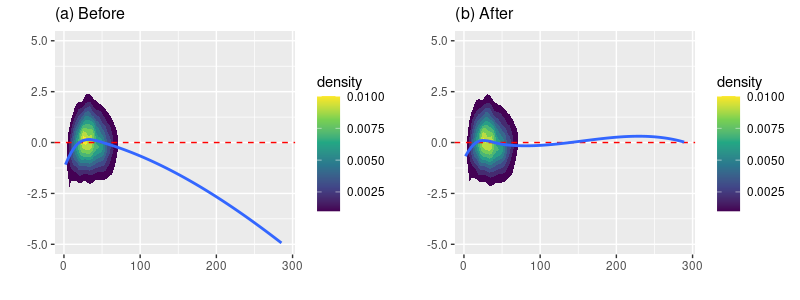}
\caption{Plots of functional residuals versus \textit{free.sulfur.dioxide} before and after adding its quadratic term to the model.}
\label{fig:winequad}
\end{figure}

(\textit{Detecting outliers}) From the diagnostic plots in the upper row of Figure \ref{fig:firstheatmap}, we observe the presence of outliers in variables  (a) {\it fixed.acidity}, (b) {\it residual.sugar}, and (c) {\it density}. These outliers drag  the LOWESS curves farther away from the center horizontal line. The detected outliers are confirmed in Figure \ref{fig:boxplot}(a)-(c). For example, in Figure \ref{fig:boxplot}(b), the highest value of {\it residual.sugar} is 65.8, which is quite distant from the second highest value 31.6. We emphasize, however, that being an outlier in its marginal distribution as seen in Figure \ref{fig:boxplot} does not necessarily suggest dropping it from the model. What really matters is its impact on the model fitting, which can be assessed through the lens of our proposed functional-residual-\textit{vs}-corvariate plots. We follow this argument and decide to remove 5 outliers labeled with the cross ($\times $) symbol. After the removal, the updated LOWESS curves in the lower row of Figure \ref{fig:firstheatmap}(a*)-(c*) are moving much closer to the center horizontal line. Nevertheless, we keep the outlier sitting on the far right of Figure \ref{fig:boxplot}(d). The reason is that it does not show to have a negative impact on the model fitting (see Figure \ref{fig:winequad}(b)) as long as we make a further adjustment of the model, which will be discussed below.

(\textit{Adding a quadratic term of free.sulfur.dioxide}) The parabola-shaped LOWESS curve in Figure \ref{fig:winequad}(a) indicates the missing of a higher order term of \textit{free.sulfur.dioxide} in the model. Figure \ref{fig:winequad}(b) shows that adding a quadratic term of this variable results in the LOWESS curve being flattened. The {\it t}-test also confirms the added quadratic term is significant at $\alpha=0.001$ level (see Table \ref{table1wine}). Using the AIC criterion, we also see a notable improvement in this final model. With the added term, the AIC value decreases from 10998 to 10813, which may imply an elevated predictive power.

\subsection{Diagnostics for bike sharing modeling}
In recent years, bike-sharing business has become increasingly popular in big cities as a convenient transportation alternative. This relatively new business has potential to decrease the dependence on automobile and reduce greenhouse gas emissions \citep{contardo2012balancing}. Around the globe, there are more than 500 bike-sharing programs which deploy over 500 thousand bicycles per year. Residents and travelers can easily rent a bike from and return it to designated stations using self service \citep{bikesharingsource}. However, existing data show that the hourly rental of bikes is highly variable, and it heavily depends on the time of a day and weather conditions. In order to reduce operational cost, it is imperative to understand what factors and how they drive the demand of bike rentals. To achieve this, we perform statistical modeling of the data from Capital Bike Sharing System at Washington D.C. \citep{bikesharingsource}. The data set contains 8734 observations of the hourly bike rentals in 2012. The explanatory variables are listed in Table \ref{table:exp_variables}.



\begin{figure}[]
\centering
\includegraphics[width =1\textwidth]{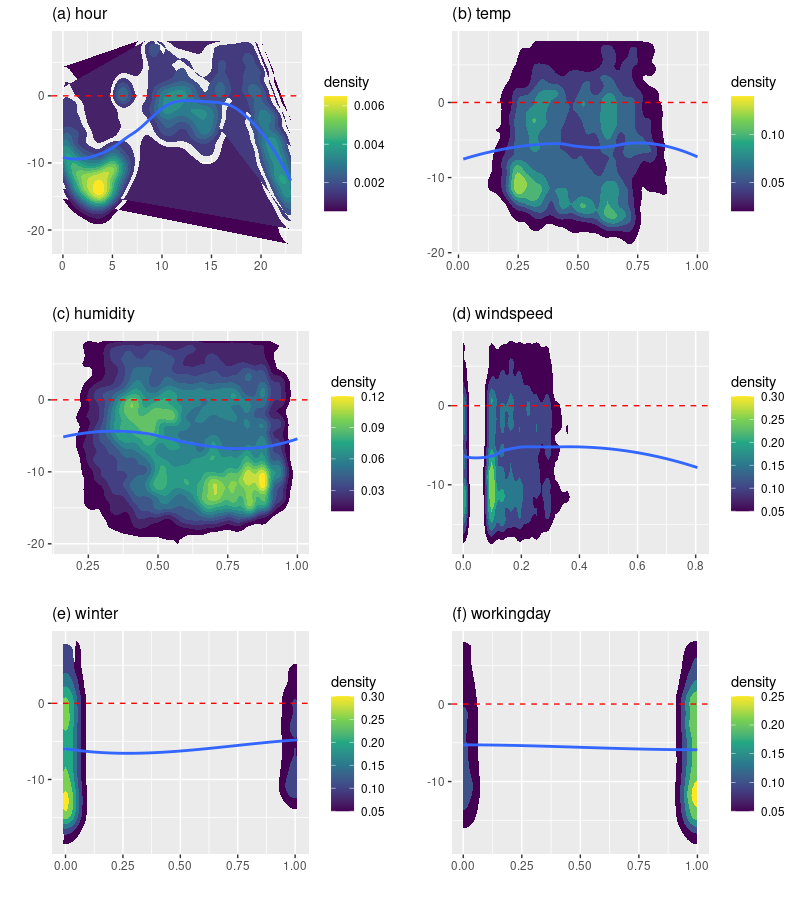}
\caption{Functional-residual-\textit{vs}-covariate plots for the initial Poisson model fitted to the Captial Bikeshare dataset.}
\label{fig:initialheat}
\end{figure}

\begin{figure}[]
\centering
\includegraphics[width =1\textwidth]{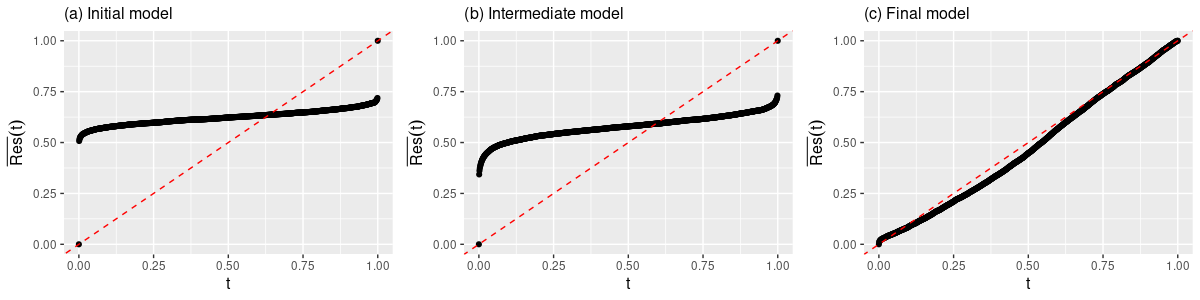}
\caption{The {\it Fn-Fn} plots for the initial, intermediate, and final models developed in the model building process.}
\label{fig:FNFN}
\end{figure}

\begin{figure}[]
\centering
\includegraphics[width =1\textwidth]{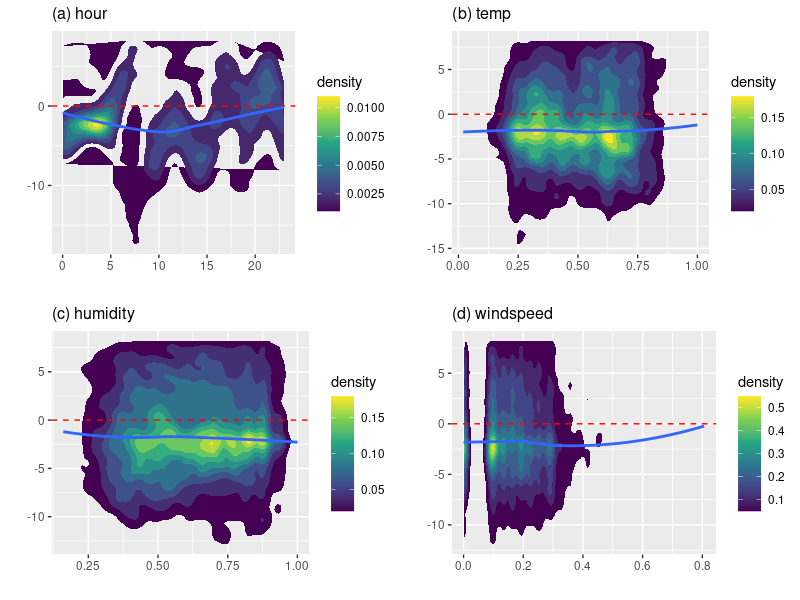}
\caption{Functional-residual-\textit{vs}-covariate plots after adding the smoothing functions of the variables {\it hour, temp, humidity,} and {\it windspeed} to the Poisson model.}
\label{fig:gampoisson}
\end{figure}

(\textit{Adding smoothing functions}) We start the modeling process by using a regular Poisson regression model and including all of the explanatory variables in the model. The model fitting result can be found in Table \ref{table:summary_models} under the title ``Initial model". To check the model fitting, we plot the functional residuals versus each of the covariates in Figure \ref{fig:initialheat}. The LOWESS curves in all the plots are falling far below the center horizontal line. This indicates that the group of small functional residuals is disproportionately large, which is also captured by heavily stretched  tails of the {\it Fn-Fn} plot in  Figure \ref{fig:FNFN}(a). In particular, the LOWESS curve in Figure \ref{fig:initialheat}(a) is neither linear nor monotonic, but instead it exhibits a pattern of strong fluctuations throughout the 24 hours in a day. This observation reflects that the factor {\it hour} is a major determinant that drives the number of rentals in a cyclical way. This cyclical influence is not captured by the linear structure in the regular Poisson model. We therefore decide to add a smoothing function to the factor {\it hour} in the model fitting. We apply the same change to the variables  {\it temp, humidity,} and {\it windspeed}, considering that they all show some degrees of fluctuations in Figure \ref{fig:initialheat}(b)-(d). These proposed changes lead us to a so-called generalized additive model. Figure \ref{fig:gampoisson} displays the updated functional-residual-\textit{vs}-covariate plots for {\it hour, temp, humidity,} and {\it windspeed}. With respect to the shape of LOWESS curves, we observe that the degrees of fluctuations have all been reduced. At the same time, the addition of smoothing functions has also helped pull all the LOWESS curves toward the center horizontal line. In what follows, we continue to show how our diagnostic tools can lead us to eventually close the remaining gap between the LOWESS curves and the center horizontal line.

\begin{figure}[]
\centering
\includegraphics[width =1\textwidth]{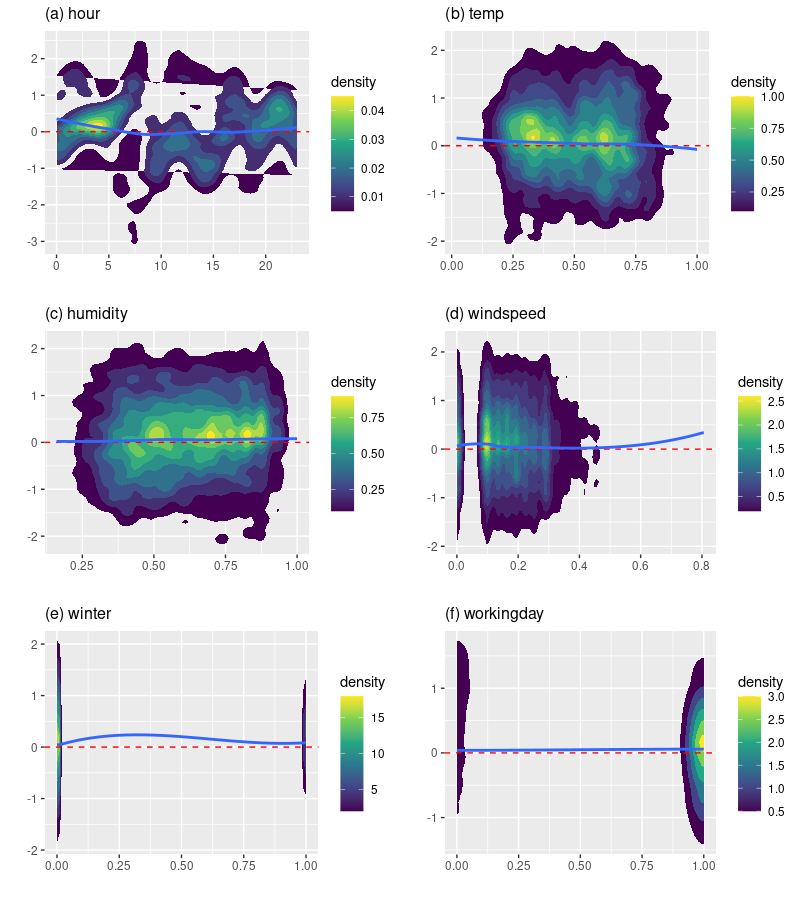}
\caption{Functional-residual-\textit{vs}-covariate plots for the final generalized additive quasi-Poisson model.}
\label{fig:finalheat}
\end{figure}

(\textit{Modeling dispersion parameter}) Although improved, the \textit{Fn-Fn} curve in Figure \ref{fig:FNFN}(b) still shows a large deviation from the 45-degree line. This is probably due to the presence of a large amount of small value data, which is also reflected in Figure \ref{fig:gampoisson} where all of the LOWESS curves are still below the center horizontal line. Note that a Poisson distribution forces its variance to be the same as the mean. However, the disproportionate amount of low bike demand (e.g., 1-5 AM) may have inflated the variance and made it much larger than the mean. Therefore, the variability of bike rentals may be underestimated in the generalized additive Poisson model. To this end, we add a dispersion parameter to the model and allow data to give an estimate instead of using the default value which is 1. We call this a generalized additive quasi-Poisson model. The model fitting result can be found in Table \ref{table:summary_models} under the title ``Final model". The estimate of the dispersion parameter is 42.998. Its considerable size confirms that the variance adjustment, suggested by our diagnostic tools, is necessary. After this update, the \textit{Fn-Fn} curve in Figure \ref{fig:FNFN}(c) turns out to be very close to the 45-degree line. This is an indication that the new model captures the inflated variability induced by very low bike demand in early AM time period. The improvement of the model is also reflected in Figure \ref{fig:finalheat}, where the LOWESS curves are now very close to the center horizontal line for the entire set of variables.

(\textit{Additional insights}) By adding smoothing functions and a dispersion parameter, the final generalized additive quasi-Poisson model has shown a remarkable improvement over the initial Poisson model. We conclude the discussion by offering an additional insight drawn from our diagnostic tools. We observe in Figure \ref{fig:finalheat}(a) that the conditional distributions of the functional residual are not homogeneous across the 24 hours. For example, the distributions (conditional on {\it hour}) in early AM hours and peak hours are very different. This heterogeneity in distribution leads to the ``residual patterns'' in Figure \ref{fig:finalheat}(a), and it is clearly not captured by the final model. The reason is that all the adjustments we have made so far (i.e., adding smoothing functions and a dispersion parameter) only addressed the issues in the first and second moments (i.e., mean and variance). To completely eliminate the remaining patterns in Figure \ref{fig:finalheat}(a), we may need to introduce  parameters to model the skewness and other higher-order moments. As \cite{li2015traffic} pointed out, a challenge of bike sharing business is the highly varying and skewed bike usage during 24 hours and across four seasons. This challenge is manifested in our diagnostic plot, which advises that input from domain experts may be needed to decide whether or not to pursue an even more complex model.

\section{A vehicle to unify point residuals}
In this section, we examine the connection of our functional residual to two point residuals developed in recent years. Our theoretical study shows that both the surrogate residual \citep{liu2018residuals, liu2020assessing} and the sign-based residual \citep{li2010test,li2012new} are point statistics that can be drawn from our functional residual. 

For ordinal outcomes, \cite{liu2018residuals} proposed a surrogate approach to defining residuals. They focused on the class of cumulative link models 
\begin{equation}
\label{eq:cumulative}
G^{-1}\left \{\Pr(Y \leq j) \right\}=\alpha_j - \bm X\bm\beta,
\end{equation}
where $G^{-1}(\cdot)$ is the link function and the intercept parameters $-\infty = \alpha_0 < \alpha_1 \cdots < \alpha_J = \infty$. The use of the logit, probit, and complementary log-log links yields the classic proportional odds, ordered probit, and proportional hazard models (see \citealp{liu2018residuals} and \citealp{liu2020assessing} for other links and models). The surrogate idea is to simulate a continuous outcome $S$ and use it as a ``surrogate'' of the ordinal outcome $Y$. They make use of the latent structure of Model~\eqref{eq:cumulative} to generate a surrogate outcome 
\[
S \sim Z \mid \alpha_{j-1} < Z \leq \alpha_j \quad \text{if}\ Y=j,
\]  
where $Z=\bm X\bm\beta + \epsilon$ ($\epsilon \sim G$) is the classical latent outcome that underlies $Y$. The residual is then defined as the difference between the surrogate outcome $S$ and its expectation, i.e., $R_{sur}=S-E(S \mid \bm X)$. We show that $R_{sur}$ is simply a derivative of our functional residual $Res(t)$ in \eqref{eq:res-func}, which nevertheless does not rely on the latent structure of the cumulative link model.

Given the data $(y, \bm x)$, as the functional residual $Res(t; y, \bm x)$ in \eqref{eq:res-func} is a distribution function, we define a {\it functional residual variable} as 
\begin{equation}
\label{eq:func-res-var}
  R_{func} \mid (y, \bm x) \sim Res(t; y, \bm x). 
\end{equation}
The following result shows that the conditional distribution of $G^{-1}(R_{func})$ is the same as that of the surrogate residual $R_{sur}$.

\begin{theorem}
\label{thm:surrogate-equiv}
 For the cumulative link model in \eqref{eq:cumulative}, our transformed functional residual variable $G^{-1}(R_{func})$ and the surrogate residual variable $R_{sur}$ follow the same conditional distribution, i.e., $G^{-1}(R_{func}) \stackrel{d}{=} R_{sur}$, given the data $(y, \bm x)$.
\end{theorem}

Theorem~\ref{thm:surrogate-equiv} suggests that the surrogate residual can be viewed as a byproduct of our functional residual. The former borrows the latent variable technique, whereas the latter does not. Instead, the functional residual applies to non-cumulative-link models, such as the adjacent-category logit model studied in Section 3. It therefore broadens the scope of applicable models and provides a unified treatment for binary, ordinal, and count data.  

The result below shows that the sign-based residual \citep{li2012new} $$R_{SBS}=E\{sign(y-Y)\}=Pr\{y<Y \}-Pr\{ y>Y\}$$can also be derived from our functional residual. 
This residual is the difference between the probabilities of the ordinal variable greater than or less than the observed value.

\begin{theorem}
Given the data $(y, \bm x)$, the sign-based residual $R_{SBS}$ can be expressed using the expectation of the functional residual variable, i.e., $R_{SBS}= 2 E \{ R_{func}| (y,\bm x) \} - 1 $.
\label{thm:trans}
\end{theorem}

There is a notable difference between the sign-based and functional residuals. As discussed in depth in \cite{liu2018residuals} and \cite{liu2020assessing}, the sign-based residual only has first-moment property for model diagnostics (i.e., $E(R_{SBS})=0$). It therefore has the same problem as Pearson/deviance residuals as observed in Figure \ref{fig:ordinal-correct}(c)-(d), i.e., it may show unusual patterns even for correctly specified models. Our functional residual solves this issue. As demonstrated both theoretically and numerically, it can be used in a similar way to the ordinary residual for linear regression models. Any pattern observed in our diagnostic plots is an indication of model misspecification, and it also provides clues for model treatment.


\section{Summary}
In this article, we have proposed a unified framework of model diagnostics for discrete data regression. The framework is based on a new notion of functional residual, which is fundamentally different from the traditional residual concept. We have developed theoretical properties of functional residuals to support the use of the residual-{\it vs}-covariate plot and {\it Fn-Fn} plot as diagnostic tools. The simulated examples in a variety of settings and case studies have shown that our framework has the power to detect misspecification of many important components of ordinal/count data regression models. It has been demonstrated that the interpretation of our diagnostic plots is similar to that of those widely used for linear regression models. Our framework, therefore, bridges the interpretation gap between models with a discrete outcome and those with a continuous outcome. This is achieved for a broad class of generalized linear models and any common type of discrete data.


\bibliographystyle{asa}
\bibliography{bib_diagnostics}

\begin{thebibliography}{31}
\newcommand{\enquote}[1]{``#1''}
\expandafter\ifx\csname natexlab\endcsname\relax\def\natexlab#1{#1}\fi

\bibitem[{Archer and Lemeshow(2006)}]{archer2006goodness}
Archer, K.~J. and Lemeshow, S. (2006), \enquote{Goodness-of-fit test for a
  logistic regression model fitted using survey sample data,} \textit{The Stata
  Journal}, 6, 97--105.

\bibitem[{Archer et~al.(2007)Archer, Lemeshow, and Hosmer}]{ARCHER20074450}
Archer, K.~J., Lemeshow, S., and Hosmer, D.~W. (2007), \enquote{Goodness-of-fit
  tests for logistic regression models when data are collected using a complex
  sampling design,} \textit{Computational Statistics and Data Analysis}, 51,
  4450--4464.

\bibitem[{Bl{\"o}chlinger and Leippold(2011)}]{blochlinger2011new}
Bl{\"o}chlinger, A. and Leippold, M. (2011), \enquote{A new goodness-of-fit
  test for event forecasting and its application to credit defaults,}
  \textit{Management Science}, 57, 487--505.

\bibitem[{Bondell(2007)}]{bondell2007testing}
Bondell, H.~D. (2007), \enquote{Testing goodness-of-fit in logistic
  case-control studies,} \textit{Biometrika}, 94, 487--495.

\bibitem[{Box(1976)}]{box1976science}
Box, G.~E. (1976), \enquote{Science and statistics,} \textit{Journal of the
  American Statistical Association}, 71, 791--799.

\bibitem[{Contardo et~al.(2012)Contardo, Morency, and
  Rousseau}]{contardo2012balancing}
Contardo, C., Morency, C., and Rousseau, L.-M. (2012), \textit{Balancing a
  Dynamic Public Bike-sharing System}, vol.~4, Cirrelt: Montreal.

\bibitem[{Cortez et~al.(2009)Cortez, Cerdeira, Almeida, Matos, and
  Reis}]{winesource}
Cortez, P., Cerdeira, A., Almeida, F., Matos, T., and Reis, J. (2009),
  \enquote{Modeling wine preferences by data mining from physicochemical
  properties,} \textit{Decision Support Systems}, 47, 547--553.

\bibitem[{De~Jong and Heller(2008)}]{de2008generalized}
De~Jong, P. and Heller, G.~Z. (2008), \textit{Generalized Linear Models for
  Insurance Data}, Cambridge University Press: NY, USA.

\bibitem[{Fagerland and Hosmer(2016)}]{fagerland2016tests}
Fagerland, M.~W. and Hosmer, D.~W. (2016), \enquote{Tests for goodness of fit
  in ordinal logistic regression models,} \textit{Journal of Statistical
  Computation and Simulation}, 86, 3398--3418.

\bibitem[{Fanaee-T and Gama(2014)}]{bikesharingsource}
Fanaee-T, H. and Gama, J. (2014), \enquote{Event labeling combining ensemble
  detectors and background knowledge,} \textit{Progress in Artificial
  Intelligence}, 2, 113--127.

\bibitem[{Faraway(2016)}]{faraway2016extending}
Faraway, J.~J. (2016), \textit{Extending the linear model with R: generalized
  linear, mixed effects and nonparametric regression models, Second Edition},
  Chapman and Hall/CRC: Florida, US.

\bibitem[{Fern{\'a}ndez and Liu(2016)}]{fernandez2016goodness}
Fern{\'a}ndez, D. and Liu, I. (2016), \enquote{A goodness-of-fit test for the
  ordered stereotype model,} \textit{Statistics in Medicine}, 35, 4660--4696.

\bibitem[{Fern{\'a}ndez et~al.(2020)Fern{\'a}ndez, Liu, Arnold, Nguyen, and
  Spiess}]{fernandez2020model}
Fern{\'a}ndez, D., Liu, I., Arnold, R., Nguyen, T., and Spiess, M. (2020),
  \enquote{Model-based goodness-of-fit tests for the ordered stereotype model,}
  \textit{Statistical Methods in Medical Research}, 29, 1527--1541.

\bibitem[{Franses and Paap(2001)}]{franses2001quantitative}
Franses, P.~H. and Paap, R. (2001), \textit{Quantitative Models in Marketing
  Research}, Cambridge University Press: Cambridge, UK.

\bibitem[{Frees(2009)}]{frees2009regression}
Frees, E.~W. (2009), \textit{Regression Modeling with Actuarial and Financial
  Applications}, Cambridge University Press: NY, USA.

\bibitem[{Hosmer et~al.(1997)Hosmer, Hosmer, Le~Cessie, and
  Lemeshow}]{hosmer1997comparison}
Hosmer, D.~W., Hosmer, T., Le~Cessie, S., and Lemeshow, S. (1997), \enquote{A
  comparison of goodness-of-fit tests for the logistic regression model,}
  \textit{Statistics in Medicine}, 16, 965--980.

\bibitem[{Lambdin(2012)}]{lambdin2012significance}
Lambdin, C. (2012), \enquote{Significance tests as sorcery: Science is
  empirical—significance tests are not,} \textit{Theory \& Psychology}, 22,
  67--90.

\bibitem[{Li and Shepherd(2010)}]{li2010test}
Li, C. and Shepherd, B. (2010), \enquote{Test of association between two
  ordinal variables while adjusting for covariates,} \textit{Journal of the
  American Statistical Association}, 105, 612--620.

\bibitem[{Li and Shepherd(2012)}]{li2012new}
--- (2012), \enquote{A new residual for ordinal outcomes,} \textit{Biometrika},
  99, 473--480.

\bibitem[{Li et~al.(2015)Li, Zheng, Zhang, and Chen}]{li2015traffic}
Li, Y., Zheng, Y., Zhang, H., and Chen, L. (2015), \enquote{Traffic Prediction
  in a Bike-Sharing System,} in \textit{Proceedings of the 23rd ACM
  International Conference on Advances in Geographical Information Systems},
  ACM SIGSPATIAL 2015.

\bibitem[{Lipsitz et~al.(1996)Lipsitz, Fitzmaurice, and
  Molenberghs}]{lipsitz1996goodness}
Lipsitz, S.~R., Fitzmaurice, G.~M., and Molenberghs, G. (1996),
  \enquote{Goodness-of-fit tests for ordinal response regression models,}
  \textit{Journal of the Royal Statistical Society: Series C (Applied
  Statistics)}, 45, 175--190.

\bibitem[{Liu et~al.(2021)Liu, Li, Yu, and Moustaki}]{liu2020assessing}
Liu, D., Li, S., Yu, Y., and Moustaki, I. (2021), \enquote{Assessing partial
  association between ordinal variables: quantification, visualization, and
  hypothesis testing,} \textit{Journal of the American Statistical
  Association}, 116, 955--968.

\bibitem[{Liu and Zhang(2018)}]{liu2018residuals}
Liu, D. and Zhang, H. (2018), \enquote{Residuals and diagnostics for ordinal
  regression models: a surrogate approach,} \textit{Journal of the American
  Statistical Association}, 113, 845--854.

\bibitem[{Nattino et~al.(2020)Nattino, Pennell, and
  Lemeshow}]{nattino2020assessing}
Nattino, G., Pennell, M.~L., and Lemeshow, S. (2020), \enquote{Assessing the
  goodness of fit of logistic regression models in large samples: A
  modification of the Hosmer-Lemeshow test,} \textit{Biometrics}, 76, 549--560.

\bibitem[{Nuzzo(2014)}]{nuzzo2014statistical}
Nuzzo, R. (2014), \enquote{Statistical errors,} \textit{Nature}, 506, 150.

\bibitem[{Pyne(1979)}]{pyne1979single}
Pyne, D.~A. (1979), \enquote{Single-variable Poisson regression: A
  goodness-of-fit test and the comparison of regression coefficients,}
  \textit{Journal of the American Statistical Association}, 74, 489--493.

\bibitem[{Simmons et~al.(2011)Simmons, Nelson, and
  Simonsohn}]{simmons2011false}
Simmons, J.~P., Nelson, L.~D., and Simonsohn, U. (2011),
  \enquote{False-positive psychology: Undisclosed flexibility in data
  collection and analysis allows presenting anything as significant,}
  \textit{Psychological science}, 22, 1359--1366.

\bibitem[{Stokes et~al.(2012)Stokes, Davis, and Koch}]{stokes2012categorical}
Stokes, M., Davis, C., and Koch, G. (2012), \textit{Categorical Data Analysis
  Using SAS, Third Edition}, SAS Institute: Cary, NC, USA.

\bibitem[{Tutz(2012)}]{tutz2012regression}
Tutz, G. (2012), \textit{Regression for Categorical Data}, Cambridge University
  Press: Cambridge, UK.

\bibitem[{Wang and Famoye(1997)}]{wang1997modeling}
Wang, W. and Famoye, F. (1997), \enquote{Modeling household fertility decisions
  with generalized Poisson regression,} \textit{Journal of Population
  Economics}, 10, 273--283.

\bibitem[{Wasserstein and Lazar(2016)}]{wasserstein2016asa}
Wasserstein, R.~L. and Lazar, N.~A. (2016), \enquote{The ASA Statement on
  p-Values: Context, Process, and Purpose,} \textit{The American Statistician},
  70, 129--133.

\end{thebibliography}

\newpage

\section*{Appendix A. Additional examples for section 3.2}
\begin{figure}[]
\centering
\includegraphics[width =1\textwidth]{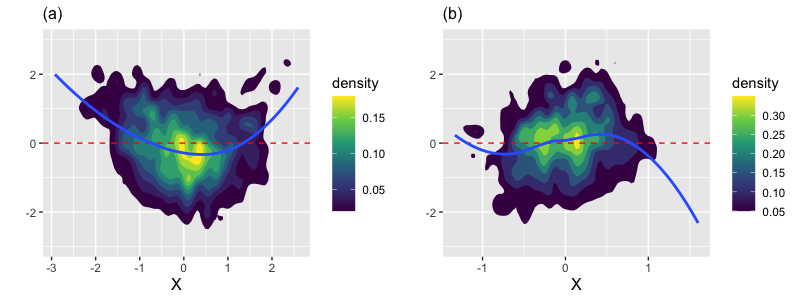}
\caption{Functional-residual-\textit{vs}-covariate plots when the quadratic term $X^2$ is missing in the working model (left) and cubic term $X^3$ is missing in the working model (right) for the count data (Example \ref{exp:poi2}). }
\label{fig:highorder}
\end{figure}

\begin{example}[Missing of a higher order term]
\label{exp:poi2}
Suppose that the data are generated from the same Poisson model used in Example \ref{exp:poi1}, whereas the quadratic term $X^2$ is not included in the working model. Our functional-residual-$vs$-covariate relationship is displayed in Figure \ref{fig:highorder}(a). The U-shape of Lowess curve unequivocally indicates the missing of a quadratic term in the working model. It is worth noting that Figure~\ref{fig:highorder}(a) here is similar to Figure~\ref{fig:ordinal-quadratic}(b) where a quadratic form is missing in modeling ordinal data. We stress that this similarity across different types of data is not a coincidence but an appealing feature consistently seen throughout the following examples.

As a further demonstration, we simulate 1000 data points $(x_i,y_i)$ from the model with a cubic term $log(E(Y|X))=\alpha+\beta_1  X+ \beta_2 X^2+\beta_3 X^3,$ where $X \sim \mathcal{N}(0,0.5)$ and $(\alpha, \beta_1, \beta_2, \beta_3)=(0.8, -0.2, 0.5, -0.5)$. When the $X^3$ term is not included in the working model, the LOWESS curve in Figure~\ref{fig:highorder}(b) shows a cubic pattern, which indicates the missing of an $X^3$ term in the model. 

\end{example}

\begin{figure}[]
\centering
\includegraphics[width =1\textwidth]{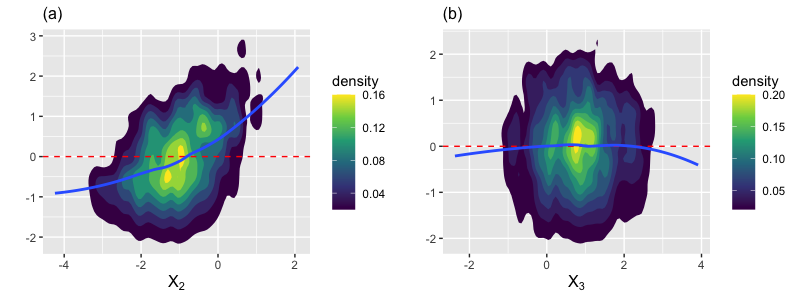}
\caption{Functional-residual-\textit{vs}-covariate plots when a covariate $X_2$ (left) is correlated with the count response $Y$ whereas $X_3$ (right) is not in the setting of Example~\ref{exp:poicov}.}
\label{fig:covarpoi}
\end{figure}

\begin{example}[Missing of relevant covariates]
\label{exp:poicov}
We simulate 1000 samples from a Poisson model with multiple covariates $log(E(Y|X))=\alpha+\beta_1  X_1+ \beta_2 X_2 +\beta_3 X_3$, where $X_1 \sim  \mathcal{N}(0,0.8)$, $X_2 \sim  \mathcal{N}(-1,1)$, and $X_3 \sim \mathcal{N}(0.8, 0.9)$. We set the coefficients $(\alpha, \beta_1, \beta_2, \beta_3)=(0.5, 0.25, 0.5, 0)$ such that both $X_1$ and $X_2$ are correlated with the count outcome $Y$ while $X_3$ is not related as $\beta_3=0$. We fit a Poisson model to the simulated data using $X_1$ only. We investigate the roles of $X_2$ and $X_3$, which are not modeled, through the lens of the functional-residual-\textit{vs}-covariate plots presented in Figure \ref{fig:covarpoi}. We observe in Figure \ref{fig:covarpoi}(a) a positive relationship between our functional residuals and $X_2$, which indicates that $X_2$ may help explain a significant proportion of the residual variability. To the contrary, the symmetrically distributed ``heat" in Figure \ref{fig:covarpoi}(b) does not suggest any correlation between our functional residuals and $X_3$. This observation is in line with the way we set $X_3$ as independent of $Y$.

\end{example}

\begin{figure}[]
\centering
\includegraphics[width =1\textwidth]{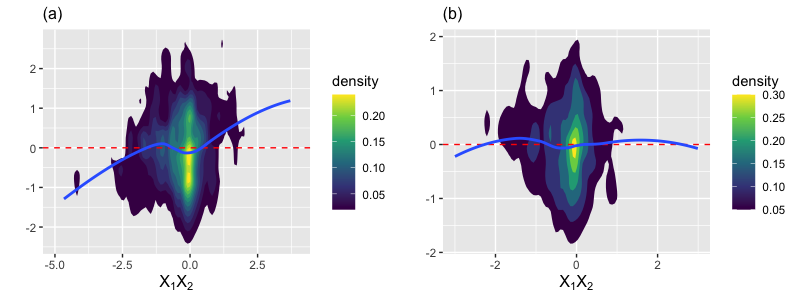}
\caption{Functional-residual-\textit{vs}-covariate plots before (left) and after (right) the interaction term $X_1 X_2$ is included in the Poisson model for the count data (Example \ref{exp:poiint}).}
\label{fig:interactionpoi}
\end{figure}

\begin{example}[Missing of an interaction]
\label{exp:poiint}
We simulate 1000 samples from a Poisson model with an interaction term $log(E(Y|X))=\alpha+\beta_1  X_1+ \beta_2 X_2+\beta_3 X_1 X_2$, where $X_1 \sim  \mathcal{N}(0.5,1)$, $X_2 \sim  \mathcal{N}(-1,0.7)$, and $(\alpha, \beta_1, \beta_2, \beta_3)=(-0.1, 0.8, -0.5, 0.6)$. We fit a Poisson model without the interaction term $X_1 X_2$. To see if our approach can detect the missing of interactions, we plot our functional residuals versus $X_1X_2$ in Figure \ref{fig:interactionpoi}(a). The LOWESS curve exhibits a clear increasing pattern compared to the plot in Figure \ref{fig:interactionpoi}(b) where the model is specified correctly including the interaction term.

\end{example}
\newpage
\section*{Appendix B. Proofs of the theoretical results}
\subsection*{Proofs for Section 2.3}

\begin{proof}[Proof of Theorem \ref{thm:conditional-exp}]
Let $y_{max}$ be the maximum of all the possible values of $Y$ ($y_{max}=\infty$ for count data), and $y_t$ be such that $\pi(y_t-1;\bm x) \leq t \leq \pi(y_t; \bm x)$. Then, we can write 
\begin{eqnarray*}
E_{(Y | \bm X=\bm x)} Res(t; Y, \bm x) & = & \sum_{y=0}^{y_{max}} Res(t; y, \bm x) \Pr\{Y=y\}\\
& = & \sum_{y=0}^{y_{max}} \Pr\{ U(\pi(y-1;\bm x), \pi(y;\bm x)) \leq t \} \Pr\{Y=y\}\\
& = & \sum_{y=0}^{y_t-1} \Pr\{Y=y\} + \Pr\{ U(\pi(y_t-1;\bm x), \pi(y_t;\bm x)) \leq t \} \Pr\{Y=y_t\}\\
& = & \pi_0(y_t-1;\bm x) + \frac{t-\pi(y_t-1;\bm x)}{\pi(y_t;\bm x)-\pi(y_t-1;\bm x)}\big\{\pi_0(y_t;\bm x)-\pi_0(y_t-1;\bm x) \big\}\\
& = & t.
\end{eqnarray*}
The last equation is a result of the assumption $\pi(\cdot; \bm x) \equiv \pi_0(\cdot; \bm x)$. This completes the proof.
\end{proof}

\begin{proof}[Proof of Theorem \ref{thm:unconditional-exp}]

By the law of iterated expectations, we have
\begin{eqnarray*}
E_{(Y,\bm X)} Res(t; Y, \bm x) & = & E_{\bm X} \{E_{(Y | \bm X=\bm x)} Res(t; Y, \bm x) \} \\
& = & \int E_{(Y | \bm X=\bm x)} Res(t; Y, \bm x) \, d F_{\bm X}(\bm x) =\int t  \, d F_{\bm X}(\bm x) = t \\
\end{eqnarray*}
The establishment above holds for every $t \in [0,1]$. This completes the proof.  
\end{proof}

\begin{proof}[Proof of Corollary \ref{cor:unconditional-ave}]
To prove Corollary \ref{cor:unconditional-ave}, we only need to verify the finite variance assumption of the strong law of large numbers. The expression for the variance can be written as follows:
\begin{eqnarray*}
Var_{(Y, \bm X)} Res(t; Y, \bm X) & = & E_{(Y, \bm X)} Res(t; Y, \bm X)^2- \{ E_{(Y, \bm X)} Res(t; Y, \bm X)\}^2. \\
\end{eqnarray*}
Since $Res(t; Y, \bm X) \in [0,1]$, we immediately have $E_{(Y, \bm X)} Res(t; Y, \bm X)^2 \leq 1$. Therefore, 
$$Var_{(Y, \bm X)} Res(t; Y, \bm X) \leq 1- t^2 \leq 1,$$
which proves the variance of $Res(t; Y, \bm X)$ is finite.
\end{proof}

\begin{proof}[Proof of Theorem \ref{the:wlln}]

Since $(Y_1,\bm X_1), (Y_2,\bm X_2), \ldots$ is an infinite sequence of i.i.d. random variables, and we have that
\begin{eqnarray*}
Var(\frac{1}{n}\sum_{i=1}^n Res(t;Y_i,\bm X_i))&=&\frac{1}{n^2}Var(Res(t; Y_1, \bm X_1)+\cdots +Res(t; Y_n, \bm X_n))\\
&=&\frac{n Var(Res(t; Y, \bm X))}{n^2} \leq \frac{1}{n} .
\end{eqnarray*}
The last inequality is due to the fact that $Var(Res(t; Y_n, \bm X_n)) \leq 1$ as proved in Theorem \ref{thm:unconditional-exp}. Therefore, we can use Chebyshev's inequality on $\frac{1}{n}\sum_{i=1}^n Res(t;Y_i,\bm X_i)$, which results in
\begin{eqnarray*}
P(|\frac{1}{n}\sum_{i=1}^n Res(t;Y_i,\bm X_i)-t| \ge \epsilon) \leq \frac{Var(\frac{1}{n}\sum_{i=1}^n Res(t;Y_i,\bm X_i))}{\epsilon^2} \leq \frac{1}{n \epsilon^2}.
\end{eqnarray*}
The inequality above is equivalent to
\begin{eqnarray*}
P(|\overline{Res}(t;Y_n,\bm X_n)-t| < \epsilon) \ge 1-\frac{1}{n \epsilon^2}.
\end{eqnarray*}
As the right side of the inequality does not depend on $t$, we have
\begin{eqnarray*}
 \inf_{t \in [0,1]}P(|\overline{Res}(t;Y_n,\bm X_n)-t| < \epsilon) \xrightarrow{} 1 , \quad\text{as } n \xrightarrow{} \infty.
\end{eqnarray*}
This completes the proof.
\end{proof}

\subsection*{Proofs for Section 5}

\begin{proof}[Proof of Theorem \ref{thm:surrogate-equiv}]

First, by the definitions of latent variable and probability density function of truncated distribution, we can get the probability density function of the surrogate variable:
\begin{equation}
f_{R_s}(r_s|Y=y)=\frac{g(r_s)}{G(\alpha_{y}- \bm X \beta)-G(\alpha_{y-1}- \bm X \beta)},
\label{eq:r_s_PDF}
\end{equation}
where $g(\cdot)=G'(\cdot)$ is the probability density function(PDF). 

Secondly, the functional residual defined in (\ref{eq:func-res-var}) follows $F_{U(\pi(y-1; \bm x),\pi(y; \bm x))}(t).$
Therefore, the PDF of $R_f$ can be written as:
\begin{equation*}
f_{R_f}(r_f)=\frac{1}{\pi(y; \bm x) - \pi(y-1; \bm x)}=\frac{1}{G(\alpha_{y}- \bm X \beta)-G(\alpha_{y-1}- \bm X \beta)}
\end{equation*}
Now we define a random variable $R_H=G^{-1}(R_f)$, the PDF of $R_H$ is:
\begin{align}
f_{R_H}(r_H) & = f_{R_f}(G(r_H))  \nonumber\\ 
&=\frac{|G(r_H)|'}{G(\alpha_{y}- \bm X \beta)-G(\alpha_{y-1}- \bm X \beta)}\nonumber\\
&=\frac{g(r_H)}{G(\alpha_{y}- \bm X \beta)-G(\alpha_{y-1}- \bm X \beta)} \label{eq:r_H_PDF}
\end{align}
The comparison between (\ref{eq:r_s_PDF}) and (\ref{eq:r_H_PDF}) suggests that $G^{-1}(R_f)$ and $R_s$ follow the same distribution which completes the proof.
\end{proof}

\begin{proof}[Proof of Theorem \ref{thm:trans}]
First, we write the equation of $R_{SBS}$ based on $F_a(\cdot)$:
\begin{align*}
R_{SBS}  &= Pr\{ y>Y\}-Pr\{y<Y \} \\
&= Pr\{ Y<y\} - Pr\{Y>y \} \\
&= Pr\{ Y\leq y-1 \} - (1-Pr\{Y \leq y \}) \\
&=  (F_a(y)+F_a(y-1)) -1 
\end{align*}
Secondly, since $R_{func} \sim Res(t;y, \bm x)= F_{U(\pi(y-1; \bm x),\pi(y; \bm x))}(t)$, the conditional expectation of $E\{R_{func} | (y,\bm x)\}$ is $\frac{1}{2}(F_a(y)+F_a(y-1))$. Therefore, $2 E \{ R_{func}| (y,\bm x) \} - 1=(F_a(y)+F_a(y-1)) -1 = R_{SBS}$. This completes the proof.
\end{proof}

\section*{Appendix C. Figures}

\clearpage

\clearpage
\clearpage

\clearpage

\clearpage

\clearpage

\clearpage

\clearpage

\clearpage

\clearpage

\clearpage

\clearpage

\clearpage

\clearpage

\clearpage

\clearpage

\clearpage
\clearpage

\section*{Appendix D. Tables}

\begin{table}[h]
\centering
\caption{Summary of the initial and final models for the wine quality data.}
\label{table1wine}
\scriptsize
\begin{tabular*}{\textwidth}{l@{\extracolsep{\fill}}*{8}{c}}
\toprule
\multicolumn{1}{c}{\multirow{2}{*}{Variable Name}} & \multicolumn{3}{c}{Initial model}       &  & \multicolumn{3}{c}{Final model}                \\ \cmidrule{2-4} \cmidrule{6-8} 
\multicolumn{1}{c}{}                               & Estimate & Std.error & P-value          &  & Estimate & Std.error        & P-value          \\ \hline
volatile.acidity                                   & 3.458    & 0.210     & \textless{}0.001 &  & 3.440    & 0.216            & \textless{}0.001 \\
alcohol                                            & -0.328   & 0.049     & \textless{}0.001 &  & -0.184   & 0.054            & \textless{}0.001 \\
sulphates                                          & -1.121   & 0.178     & \textless{}0.001 &  & -1.362   & 0.184            & \textless{}0.001 \\
fixed.acidity                                      & -0.136   & 0.039     & \textless{}0.001 &  & -0.246   & 0.043            & \textless{}0.001 \\
residual.sugar                                     & -0.152   & 0.014     & \textless{}0.001 &  & -0.188   & 0.016            & \textless{}0.001 \\
free.sulfur.dioxide                                & -0.006   & 0.001     & \textless{}0.001 &  & -0.052   & 0.004            & \textless{}0.001 \\
pH                                                 & -1.295   & 0.193     & \textless{}0.001 &  & -1.647   & 0.206            & \textless{}0.001 \\
density                                            & 284.5    & 37.38     & \textless{}0.001 &  & 417.4    & 42.13            & \textless{}0.001 \\
free.sulfur.dioxide\textasciicircum{}2    &-         &     -     &    -       &              & 0.0005   & \textless{}0.001 & \textless{}0.001\\

\bottomrule
\end{tabular*}
\end{table}

\clearpage
\begin{table}[]
\center
\caption{The explanatory variables in the bike sharing data.}
\label{table:exp_variables}
\scriptsize
\begin{tabular*}{\textwidth}{@{\extracolsep{\fill}}*{3}{l}}
\toprule
Variable Name            & Possible Values                 & \multicolumn{1}{c}{Description}                                     \\ \hline
hour                     & \{0, 1, ... , 23\}              & Time in 24-hour format.                                             \\

temp & {[}0, 1{]} & The normalized temperature in Celsius, i.e., (t-tmin)/(tmax-tmin), tmin=-8, tmax=39. \\
humidity                 & {[}0, 1{]}                      & Normalized humidity. The values are derived via h/hmax , hmax =100. \\
windspeed                & {[}0, 1{]}                      & Normalized wind speed. The values are derived via w/wmax, wmax=67.  \\ 
winter                   & \{0, 1\}                        & 1: winter; 0: otherwise.                                            \\
workingday               & \{0, 1\}                        & 1: neither weekend nor holiday; 0: otherwise.                       \\
\multirow{5}{*}{weather} & \multirow{5}{*}{\{1, 2, 3, 4\}} & Weather situation:                                                  \\
       &                                 & 1: clear, few clouds, partly cloudy.                                \\
                         &                                 & 2: mist, mist + few clouds, mist + broken clouds, mist + cloudy.    \\
     &            & 3: light snow, light Rain + scattered clouds, light Rain + thunderstorm + scattered clouds.               \\
                         &                                 & 4: snow + fog, heavy rain + ice pallets + thunderstorm + mist.      \\
\hline

\end{tabular*}
\end{table}

\clearpage
\begin{table}

\caption{Summary of the initial and final models for the bike sharing data.}
\vspace{0.1in}
\label{table:summary_models}
\scriptsize
\begin{tabular*}{\textwidth}{l@{\extracolsep{\fill}}*{8}{c}}
\toprule
\textbf{}            & \multicolumn{3}{c}{Initial model}            &  & \multicolumn{3}{c}{Final model}        \\ \cmidrule{2-4} \cmidrule{6-8} 
Variable Name        & Estimate & Std.error       & P-value         &  & Estimate & Std.error & P-value         \\ \midrule
Intercept            & 4.848    & 0.005           & \textless 0.001 &  & 5.202    & 0.017     & \textless 0.001 \\
winter               & -0.323   & 0.002           & \textless 0.001 &  & -0.284   & 0.016     & \textless 0.001 \\
workingday           & 0.061    & 0.002           & \textless 0.001 &  & 0.056    & 0.010     & \textless 0.001 \\
weather              & -0.025   & 0.001           & \textless 0.001 &  & -0.139   & 0.010     & \textless 0.001 \\ \cmidrule{6-8} 
\multicolumn{4}{c}{}                                                &  & Edf      & Ref.df    & P-value         \\ \cmidrule{6-8} 
hour                 & 0.046    & \textless 0.001 & \textless 0.001 &  & 8.986    & 9.000     & \textless 0.001 \\
temp                 & 1.157    & 0.005           & \textless 0.001 &  & 4.613    & 5.632     & \textless 0.001 \\
humidity             & -1.005   & 0.005           & \textless 0.001 &  & 8.240    & 8.835     & \textless 0.001 \\
windspeed            & 0.223    & 0.006           & \textless 0.001 &  & 3.582    & 4.409     & \textless 0.001 \\

\midrule

Dispersion Parameter & \multicolumn{3}{c}{1.000}                        &  & \multicolumn{3}{c}{42.998}             \\ 
\bottomrule
\end{tabular*}%
\end{table}

\end{document}